\documentclass[12pt]{article}
\usepackage{epsf}

\usepackage{epsfig,graphics}
\usepackage {graphicx}
\usepackage {epsfig}
\usepackage {subfigure}
\usepackage {tabularx} 
\usepackage{rotate}	
\usepackage{slashed}
\usepackage{float}
\usepackage{amsmath}
\usepackage{amsfonts}
\usepackage{amssymb}
\usepackage{graphicx}
\usepackage{cite}

\usepackage{fancyhdr}
\usepackage{hyperref}

\newcommand{\bmat}{\left(\begin{array}}
\newcommand{\emat}{\end{array}\right)}

\def\yzero{\smash{\hbox{$y\kern-4pt\raise1pt\hbox{${}^\circ$}$}}}

\def\beq{\begin{equation}}
\def\eeq{\end{equation}}
\def\beqa{\begin{eqnarray}}
\def\eeqa{\end{eqnarray}}

\def\-{\hphantom{-}}
\def\ov{\overline}
\def\s2{\frac{1}{\sqrt2}}

\def\beq{\begin{equation}}
\def\eeq{\end{equation}}
\def\beqa{\begin{eqnarray}}
\def\eeqa{\end{eqnarray}}
\def\tr{{\rm tr \,}}

\def\IF{\relax{\rm I\kern-.18em F}}
\def\II{\relax{\rm I\kern-.18em I}}
\def\IP{\relax{\rm I\kern-.18em P}}
\def\IC{\relax\hbox{\kern.25em$\inbar\kern-.3em{\rm C}$}}
\def\IR{\relax{\rm I\kern-.18em R}}

\def\Dsl{\,\raise.15ex\hbox{/}\mkern-13.5mu D} 
\def\IZ{Z\kern-.4em  Z}

\def\IC{{\bf C}}
\def\IS{{\bf S}}
\def\IR{{\bf R}}
\def\IZ{{\bf Z}}

\def\IT{{\bf T}}
\def\IP{\bf P}

\def\NN{{\cal N}}

\catcode`\@=11   
\newdimen\@rotdimen
\newbox\@rotbox  

\def\@vspec#1{\special{ps:#1}}
\def\@rotstart#1{\@vspec{gsave currentpoint currentpoint translate
   #1 neg exch neg exch translate}}
\def\@rotfinish{\@vspec{currentpoint grestore moveto}}
\def\@rotr#1{\@rotdimen=\ht#1\advance\@rotdimen by\dp#1%
   \hbox to\@rotdimen{\hskip\ht#1\vbox to\wd#1{\@rotstart{90 rotate}%
   \box#1\vss}\hss}\@rotfinish}
\def\@rotl#1{\@rotdimen=\ht#1\advance\@rotdimen by\dp#1%
   \hbox to\@rotdimen{\vbox to\wd#1{\vskip\wd#1\@rotstart{270 rotate}%
   \box#1\vss}\hss}\@rotfinish}%
\def\@rotu#1{\@rotdimen=\ht#1\advance\@rotdimen by\dp#1%
   \hbox to\wd#1{\hskip\wd#1\vbox to\@rotdimen{\vskip\@rotdimen
   \@rotstart{-1 dup scale}\box#1\vss}\hss}\@rotfinish}%
\def\@rotf#1{\hbox to\wd#1{\hskip\wd#1\@rotstart{-1 1 scale}%
   \box#1\hss}\@rotfinish}%
\def\rotate{\@ifnextchar[{\@rotate}{\@rotate[l]}}
\def\@rotate[#1]#2{\setbox\@rotbox=\hbox{#2}\@nameuse{@rot#1}\@rotbox}

\catcode`\@=12

\topmargin
-1.5cm
\textwidth
15.5cm
\textheight
23.5cm
\oddsidemargin
0.7cm
\evensidemargin
0.7cm

\begin{document}

\makeatletter
\@addtoreset{equation}{section}
\makeatother
\renewcommand{\theequation}{\thesection.\arabic{equation}}
\pagestyle{empty}
\vspace{-0.2cm}
\rightline{ IFT-UAM/CSIC-18-123}
\vspace{0.5cm}
\begin{center}


\LARGE{\bf Modular Symmetries \\ and the Swampland Conjectures\\ [13mm]}

  \large{E. Gonzalo$^{1,2}$, L.E. Ib\'a\~nez$^{1,2}$ and A.M. Uranga$^1$\\[5mm]}
\small{
  ${}^{1}$ Instituto de F\'{\i}sica Te\'orica IFT-UAM/CSIC,\\[-0.3em] 
C/ Nicol\'as Cabrera 13-15, 
Campus de Cantoblanco, 28049 Madrid, Spain \\ 
${}^2$ Departamento de F\'{\i}sica Te\'orica, Facultad de Ciencias\\[-0.3em] 
Universidad Aut\'onoma de Madrid, 28049 Madrid, Spain\\[5mm]}
\small{\bf Abstract} \\[6mm]
\end{center}
\begin{center}
\begin{minipage}[h]{15.22cm}

{\footnotesize Recent string theory tests of swampland ideas  like the distance or the dS conjectures 
have been performed at weak coupling. 
Testing these ideas beyond the weak coupling regime remains challenging. We propose to exploit
the modular symmetries of the moduli effective action to check swampland constraints 
beyond perturbation theory. As an example we study the case of heterotic 4d $\NN=1$ compactifications, whose non-perturbative effective action is known to be invariant under modular symmetries acting on the K\"ahler and complex structure moduli, in particular $SL(2,\IZ)$ T-dualities (or subgroups thereof) for 4d heterotic or orbifold compactifications. Remarkably, in models with non-perturbative superpotentials, the corresponding duality invariant potentials diverge at points at infinite distance in moduli space. 
The divergence relates to towers of states becoming  light, in agreement with the distance conjecture.
We discuss specific examples of this behavior based on gaugino condensation in heterotic orbifolds. 
We show that these examples are dual to compactifications of type I' or Horava-Witten theory, in which the $SL(2,\IZ)$ acts on the complex structure of an underlying 2-torus, and the tower of light states correspond to D0-branes or M-theory KK modes.
The non-perturbative examples explored point to potentials not leading to weak coupling at infinite distance, 
but rather  diverging  in the asymptotic corners of moduli space, dynamically forbidding the access to points with global symmetries.
We perform a study of general modular invariant potentials and find  that there are dS maxima and saddle points but no
dS minima, and that all examples explored obey the refined dS conjecture. 

}

\end{minipage}
\end{center}
\newpage
\setcounter{page}{1}
\pagestyle{plain}
\renewcommand{\thefootnote}{\arabic{footnote}}
\setcounter{footnote}{0}


\tableofcontents

\section{Introduction}

There has recently been a lot of interest on the concept of  the Swampland constraints on effective theories
 \cite{swampland,WGC,OV1,OV,Obied:2018sgi,Agrawal:2018own,Ooguri:2018wrx}, see \cite{vafafederico} for a review. These are powerful conditions excluding effective theories which cannot arise in consistent theories of quantum gravity. Some of the most interesting swampland conjectures refer to the properties of the moduli space of scalars in consistent theories of quantum gravity 
 \cite{irene,emergence,Lee:2018urn,Lee:2018spm,Hebecker:2018vxz,Palti:2017elp,Lust:2017wrl,Denef:2018etk,Grimm:2018cpv}. The main list of such constraints include the following:
\begin{itemize}

\item The moduli space of scalars is non-compact.

\item  Distance conjecture: Consider a point in moduli space $p_0\in {\cal M}$. Then in the limit of
infinite distance $d(p_0,p)=t\rightarrow \infty $, an infinite tower of states appears in the effective field theory 
with exponentially decreasing masses $m\simeq e^{-\alpha t}$.

\item {\it Refined} dS conjecture \cite{Ooguri:2018wrx} (see also \cite{Garg:2018reu}): Any scalar potential $V(\phi)$ in a consistent theory of quantum gravity
must obey either 
\beq
\left|\nabla V\right|\ \geq \ \frac {c}{M_p} V \  \quad\quad
{\rm or}\quad \quad
{\rm min}(\nabla_i\nabla_jV)\ \leq \ -\frac{c'}{M_p^2} V \ .
\eeq
\end{itemize}
There are presumably interesting interplays among these conjectures. A possible connection between the last two conditions has been pointed out in \cite{Ooguri:2018wrx}, see also \cite{Hebecker:2018vxz}; also, the connection between the distance conjecture and global symmetries has been pointed out in \cite{irene}.
The first two conjectures have been tested in a number of different string theory settings, see \cite{irene,emergence,Lee:2018urn,Lee:2018spm}
and references therein.  In this paper we will explore the relations between these constraints, in the hitherto unexplored regime of low supersymmetry and strong coupling.
For some recent papers on the Swampland and the Weak Gravity Conjecture see also
\cite{masdS,WGC1,WGC2,WGC3,moreOV} and \cite{3D2D,HS}.

The string theory tests of the distance conjecture performed so far are either perturbative or involve at least eight supersymmetries (or simple circle compactifications). These tests are also often essentially kinematic, in the sense that the structure of the towers of light states is identified but a full understanding of their effect on the effective field theory is lacking. Only with the additional constraints of theories with at least 8 supersymmetries it is possible to e.g. signal the emergence of a dual theory  \cite{irene,emergence,Lee:2018urn,Lee:2018spm}.

It would certainly be interesting to try and test these conjectures in string settings with reduced supersymmetry and also e.g.  including non-perturbative induced potentials. This seems to be out of reach.  However, in this paper we make substantial progress in this direction by proposing to exploit the modular symmetries of the effective action to check swampland ideas beyond weak coupling. Indeed, the scalar moduli space of string compactifications transforms in general under modular (or paramodular) symmetries, which are discrete infinite symmetries involving transformations of the moduli. They involve in general the K\"ahler, complex structure and complex dilaton of 4d string compactifications, and are part of the duality symmetries of string theory. The prototypical example is the $SL(2,\IZ)_T$ duality of tori and orbifolds thereof. This group is generated by  $R\rightarrow 1/R$ transformations along with discrete shift symmetries for the real  part of a K\"ahler modulus $T$. The existence of this kind of modular symmetries goes however beyond the toroidal setting, since modular symmetries arise also in (exact) moduli spaces of Calabi-Yau threefolds, like the quintic \cite{Candelas:1990rm} and fairly generically in large classes of eliptic  CY threefolds see e.g.
 \cite{Candelas:1990rm,Candelas:1994hw,Haghighat:2015qdq,Haghighat:2017bep}. These generalized modular symmetries are also generated by some discrete shift symmetry and a transformation relating large and small volumes (in fact the discrete shift symmetries and their interplay with the tower of light states near points at infinite distance in moduli space has already been explored in the $\NN=2$ setup in \cite{irene,Grimm:2018cpv}.
Modular behavior arise also in non-perturbative superpotentials for F-theory compactifications on CY fourfolds \cite{witten0,witten,curio,Grimm:2007xm,Anderson:2015yzz}.
Modular symmetries in axion and inflaton potentials have also been considered in
\cite{Conlon:2016aea,Olguin-Tejo:2018pfq}

We will focus on compactifications preserving 4d $\NN=1$ supersymmetry, in which non-perturbative superpotentials for the moduli may arise. The resulting scalar potentials are thus constrained by invariance under modular transformations.
A prototypical class of examples are 4d $\NN=1$ heterotic orbifolds, for which the modular  $SL(2,\IZ)$ symmetries  were well studied in the early 90's \cite{Ferrara:1989qb,Ferrara:1989bc,Font:1990nt,Ferrara:1990ei,Nilles:1990jv,Cvetic:1991qm,Font:1990gx,Horne:1994mi}. In this case the K\"ahler potential and  the (non-perturbative) superpotential must transform such that the full K\"ahler potential (and hence the scalar potential) is invariant under modular transformations.
In the case of a single modulus  (little is known about modular symmetries for the multimoduli case)
the superpotential must be an automorphic form with definite modular weight, which is a very restrictive condition on the theory. 

The study of the modular invariant effective actions of a single modulus in general 4d $\NN=1$ string vacua is
interesting in its own right, and was already considered  in the early 90's, see e.g. \cite{Cvetic:1991qm,Font:1990gx,Horne:1994mi}. We revisit these results from the perspective of the swampland conjectures.
One finds in particular that modular invariant potentials necessarily imply the existence of essential singularities at points at infinite distance in moduli space (e.g. large/small volume). We display explicitly that in concrete string examples the divergence arises from infinite towers of states becoming light in those limits, in agreement with swampland distance conjecture expectations. Hence, our analysis remarkably relates the swampland distance conjecture with the behavior of the moduli effective action under modular transformations. In addition the modular symmetries for the moduli require the existence of the states to provide a physical understanding of the singularities. Notice that this is a novel behavior, as compared with the previous discussions in 4d $\NN=2$ vacua; rather than the emergence of a dual description, what we find is that the infinite distance points are dynamically (exponentially) censored.

We describe different specific string contexts in which such modular invariant non-perturbative potentials appear, and discuss the relevant towers of states at infinite distance points.
We revisit modular invariant gaugino condensation in 4d $\NN=1$ heterotic toroidal  orbifolds,  already discussed in the early 90's \cite{Font:1990nt,Ferrara:1990ei,Nilles:1990jv}. In orbifolds with $\NN=2$ subsectors, there are moduli-dependent one-loop threshold corrections to the gauge couplings 
 \cite{Dixon:1990pc,Antoniadis:1991fh,Antoniadis:1992rq}, which can enter the moduli-dependent  non-perturbative superpotential. In this case the singularities appear due to towers of KK or winding states on the fixed $\IT^2$ in the $\NN=2$ sectors.  

A key observation is that the  moduli dependence relevant for invariance under modular transformations arises from subsectors with 4d $\NN=2$ supersymmetry. Their properties are thus controlled by the properties of heterotic compactifications on K3$\times \IT^2$. This observation allows us to explore the dynamics of moduli using other dual descriptions, in particular  type IIB orientifolds, type I' compactifications, and compactifications of Horava-Witten theory. These pictures allow a clear identification of the tower of states becoming light near the point at infinity in moduli space, and which are ultimately responsible for the appearance of the divergence in the scalar potential, and which correspond to D0-branes or KK momentum states in the latter pictures. This discussion enormously increases the applicability of our results on swampland conjectures from invariance under modular transformations to other string compactifications. In particular, non-perturbative superpotentials with modular behavior have been obtained in the literature in a variety of settings, see e.g. \cite{witten0,witten,curio,Grimm:2007xm,Anderson:2015yzz}. We expect that the lessons we present in this paper extend to more general string compactifications.

The fact that many string models lead to theories invariant under modular transformations, motivates the study of general effective theories of scalars with modular invariant potentials in the context of the swampland conjectures. They may be considered as non-trivial  candidate field theories which may be consistently coupled to quantum gravity. This consistency would also require fully fledged string theory to give a physical meaning to their behavior and singularities.  In this paper, we perform a detailed study of different single-modulus effective actions with non-trivial $SL(2,{\bf Z})$-invariant scalar potentials, and explore their extrema. We find AdS and Minkowski vacua, both SUSY and non-SUSY. We also find dS maxima and saddle points, but we have been unable to find a dS minimum. Our analysis show that the refined dS conjecture applies  in all examples studied (including the second condition, for dS maxima). We find in our examples no runaway behavior leading to e.g. possible quintessence dynamics.

We emphasize that in these effective theories the invariant potentials always diverge exponentially at points at infinity, so that the moduli are dynamically forbidden to access them. In their UV completion, the infinite tower of states demanded by the swampland distance conjecture should be responsible for the dynamical generation of the divergent potential, and thus for the impossibility to access the regions in which global symmetries are recovered.

The structure of the rest of this paper is as follows.  In Section \ref{sec:general} we review the structure of single-modulus $SL(2,\IZ)$-invariant scalar potentials in 4d $\NN=1$ supergravity. We show the superpotential must have modular weight $(-3)$, implying the appearance of singularities at the boundary of moduli space. In Section \ref{sec:string-examples} we discuss concrete string theory setups with non-perturbative superpotentials satisfying these properties under modular transformations, and describe the UV degrees of freedom providing the infinite towers of states inducing the divergent contributions to the potential. In section \ref{sec:gaugino} we review superpotentials from gaugino condensation in heterotic toroidal orbifolds, and subsequently in section \ref{sec:hw} we turn to other dual realizations, including models in type I' or Horava-Witten theory, which nicely isolate the towers of states dominating at infinite distances (and which correspond to D0-branes or KK momentum modes). In Section 4 we discuss effective field theories with general single-modulus $SL(2,\IZ)$-invariant scalar potential, and study the validity of the (refined) swampland conjectures to their extrema.  Finally, in section \ref{sec:conclu} we offer our final remarks. 
Appendix A collects some useful properties of modular functions and Appendix B an overview of the extrema in  models with two complex scalars.

\section{Modular symmetries and invariant potentials}
\label{sec:general}

We want to study how modular symmetries can constraint the effective action of string compactifications,
and how these constraints fit with swampland ideas.  We focus on the prototypical example of such symmetry, the modular group $SL(2,{\bf Z})$,  which is ubiquitous in string theory. In this section we review the scalar potential for 4d $\NN=1$ supergravity theories of a single modulus $T$ invariant under the 
modular group  $SL(2,{\bf Z})$, see   \cite{Cvetic:1991qm}. This modular symmetry is generated by the transformations
\beq
T\ \longrightarrow \ \frac {aT+b}{cT+d} \ ,  \ a,b,c,d\in {\bf Z}, ad-bc=1 \ .
\eeq
acting on the complex modulus $T=\theta+i t$. The two generators are $T\to T+1$, which implies a discrete periodicity for the real part $\theta$, making it  axion-like; and $T\to -1/T$, which relates small and large $t$, e.g. $t\to 1/t$ for $\delta=0$.

In general $T$ may correspond to a string modulus like
K\"ahler, complex structure or complex dilaton, depending on the examples considered, see Section \ref{sec:string-examples} for examples. For concreteness, we carry out the discussion in terms of $T$ being a K\"ahler modulus
The effective action will be determined by the full supergravity K\"ahler potential
\beq 
G(T,T^*)\ =    -\,3 \log\,(T-T^*) \,+\, \log|W|^2 \ .
\label{potone}
\eeq
This corresponds to e.g. the large K\"ahler modulus dependence of a Calabi-Yau (CY) string compactification with
$h_{11}=1$. It also corresponds to the K\"ahler potential for the overall K\"ahler modulus in a toroidal/orbifold compactifications.

Under modular transformations one has $(T-T^*)\rightarrow (T-T^*)/|cT+d|^2$, so that modular invariance of the K\"ahler potential dictates
that the superpotential should have modular weight (-3), i.e.
\beq
W(T)\ \longrightarrow \  e^{i\delta(a,b,c,d)}\, \frac {W(T)}{(cT+d)^3} \  \ ,
\eeq
where $\delta$ is a moduli-independent phase, which can depend on the $SL(2,\IZ)$ transformation.

We now revisit the general study of T-duality invariant potentials in \cite{Cvetic:1991qm}. A general and useful way to parametrize functions with given modular weight is through the Dedekind function $\eta(T)$, which has modular weight $1/2$,   $\eta(T)\rightarrow (cT+d)^{1/2}\, \eta(T)$. Hence we can always write
\beq
 W\ =\  \frac {H(T)}{\eta(T)^ 6}
 \label{pottwo}
 \eeq
with $H(T)$ being a modular invariant holomorphic function. This can always be expressed as a function of the absolute modular invariant function $j(T)$ (see Appendix A).

As explained in appendix  A, if one insists in avoiding singularities within the fundamental domain, it can be proven that $H(T)$ must be of the form
\beq
H(T)\ =\ (j(T)-1728)^{m/2}j(T)^{n/3}{\cal P}(j(T))
\label{generalH}
\eeq
where $m,n$ are positive integers and ${\cal P}$ is a polynomial on $j$. Without loss of generality
the zeros of ${\cal P}(j)$ may be chosen different from the $SL(2,{\bf Z})$ self-dual  points $T=i,e^{i2\pi/3}$.
Another equivalent way to write the same expression is
\beq
H(T)\ =\ \left( \frac {G_6(T)}{\eta(T)^{12}}\right) ^m 
 \left( \frac {G_4(T)}{\eta(T)^{12}} \right) ^n {\cal P}(j(T)) \ ,
 \label{superpogen}
 \eeq
 where $G_{4,6}$ are Eisenstein functions of weight 4 and 6 respectively, see Appendix A.
 
Note that these superpotentials with no additional singularities in the fundamental domain, necessarily diverge exponentially as $T\rightarrow i\infty,0$. Such a behavior is in principle rather surprising, since in perturbation theory potentials are known to scale like a power of $1/t$. Thus this peculiar behavior, if present, requires non-perturbative physics \footnote{This does not imply the absence of tree level superpotential couplings. These are indeed present, but necessarily involve charged matter fields, which transform non-trivial under the modular group and account for the matching of the total modular weight for the superpotential.}. In coming sections we will provide explicit string models with  this divergent behavior, including the gaugino condensation example studied in \cite{Font:1990nt}. These constructions show that the exponential singularities arise from towers of states becoming light, in nice agreement with the swampland distance conjecture.  

It is interesting to consider the simplest case with $H(T)=1$, which is indeed realized in concrete string constructions, as we will see in coming sections. The scalar potential is given by the simple expression
\beq
V\ =\ \frac {1}{8({\rm Im}\,T)^3\, |\eta(T)|^{12}}\left(\, \frac {3({\rm Im}\,T)^2}{\pi^2}\,|{\hat G}_2|^2-3\right) \ ,
\label{potential-eisenstein}
\eeq
where ${\hat G}_2$ is the non-holomorphic weight-2 Eisenstein function
\beq
{\hat G}_2(T,T^*)\ =\ G_2(T)\ -\ \frac {\pi}{{\rm Im}\,T}  \ .
\eeq
This potential (\ref{potential-eisenstein}) is explicitly modular invariant and diverges like $e^{\pi t}$ for large ${\rm Im}\, T=t\to\infty$. Extrema of this potential as well as more general potentials with arbitrary $H(T)$ will be discussed in section \ref{sec:potential}.  As we discuss in the next section, this type of potentials arise in gaugino condensation in heterotic orbifolds, and other related examples.

One may worry that the above computations can be trusted only for large ${\rm Im}\, T$, and that there may be corrections in the deep interior of moduli space. However, the key statement is that modular invariance strongly restricts the structure of the K\"ahler potential, the superpotential and thus the scalar potential for moduli. More formally, the K\"ahler potential and superpotential are not mere functions over moduli space, but rather sections of non-trivial bundles over moduli space. Corrections may change the specific form of the section, but cannot change the non-trivial topology of the bundle, which is encoded in the modular properties we are highlighting.
Hence, in short, the superpotential must be an holomorphic form of modular weight $-3$. Barring singularities at finite distance in moduli space (which would have no clear physical origin, as will be clear in the examples in coming sections) this implies the existence of singularities at ${\rm Im}\, T \rightarrow \infty$, which thus seems unavoidable.

The existence of this singular behavior as one approaches points at infinite distance in moduli space,  resonates with the swampland conjectures. For ${\rm Im}\, T\gg 1$ in Planck units,  the potential  dynamically  censors the limits in which
the continuous shift symmetries are recovered. Furthermore, the divergences in those limits stem from towers of almost massless states, in agreement with the distance swampland conjecture.

We would like to emphasize that, nevertheless, this behavior is in stark contrast with the behavior at large moduli found in other examples studied in the literature, possibly due to the extended $\NN=2$ supersymmetry on which they are based.
In those examples, e.g. \cite{irene}, the towers of states near those points typically induce corrections to the metric in moduli space, possibly making the infinite distance emergent; these states sometimes admit very explicit descriptions in terms of a dual theory \cite{Lee:2018urn,Lee:2018spm}. In the non-perturbative framework here described, access to these points is excluded not merely from kinematics in moduli space, but also from the dynamics inducing a potential. The models are thus trapped in a non-perturbative regime far from any weak coupling expansion. Perturbativity for the SM should then appear as an infrared effect, as suggested e.g.in \cite{emergence}.

\medskip

As explained, similar analysis applies when the relevant modulus correspond to other string moduli. The case in which it corresponds to the complex dilaton $S$ in heterotic compactifications was put forward in \cite{Font:1990gx}, in the first paper introducing S-duality. Although in principle $SL(2,\IZ)$ S-duality could be expected to be a symmetry only in finite theories, it was argued that it may apply to some extent in some subsector of theories with reduced  or no supersymmetry.  Indeed, modular invariant superpotentials involving complexified string coupling have appeared in the context of 4d $\NN=1$ F-theory compactifications \cite{witten,curio,Grimm:2007xm,Anderson:2015yzz}.

Following the same arguments as above the limits $S\rightarrow i\infty, 0$ would be dynamically 
forbidden, and the weak coupling limit $g\rightarrow 0$ would be censored, in agreement with absence of continuous global symmetries.
For large $S$, corresponding to $g\rightarrow 0$, the heterotic string becomes tension-less since one has for the string scale $M_s^2={M_p^2}/({8\, {\rm Im}\,S})$.
In ref.\cite{Horne:1994mi} the divergent limits in the S-dual potentials were avoided by canceling the divergence  choosing a $H(S)$
vanishing at $S\rightarrow i\infty, 0$. However in such a case, as discussed above, there are necessarily singularities in the fundamental region, with no obvious physical interpretation.

\section{String theory examples}
\label{sec:string-examples}

\subsection{Gaugino condensation in 4D heterotic string}
\label{sec:gaugino}

In this section we will discuss how  $\NN=1$ superpotentials of the type discussed in the previous section arise in string theory. In particular, we will consider  4d $\NN=1$ heterotic Abelian orbifolds. We first focus on the moduli-dependent threshold corrections for gauge couplings. Upon gaugino condensation, such threshold corrections can give rise to non-perturbative moduli-dependent superpotentials which are modular forms of the appropriate weight.

Threshold corrections to 4d gauge kinetic functions arise from a perturbative 1-loop computation. We eventually focus on Abelian orbifolds, which are exact CFTs, and in particular are interested in the moduli-dependent piece, which arise in ${\cal N}=2$ subsectors, i.e.  those associated to elements of the orbifold group leaving some $\IT^2$ fixed. In those sectors the structure is essentially that of a compactification of heterotic on  K3$\times \IT^2$, and the discussion can be phrased in this more general setting (eventual breaking to $\NN=1$ can be obtained by further orbifold twists), following the classical computation in  \cite{Dixon:1990pc,Antoniadis:1991fh,Antoniadis:1992rq}, also for comparison with the section to follow. 

From general results in $\NN=2$ theories, the gauge kinetic functions can only depend on moduli in vector multiplets. Thus, since the K3 moduli belong to hypermultiplets, they cannot appear in the threshold correction. Hence, we are interested in threshold corrections depending on the K\"ahler and complex structure moduli $T$, $U$ of the $\IT^2$, which are in vector multiplets. We will pay special attention to the action of the modular groups $SL(2,{\bf Z})_T\times SL(2,{\bf Z})_U$ on these moduli.

The thresholds corrections depending on the moduli of the $\IT^2$ for the resulting ${\cal N}=2$ theories arise from momentum and winding modes on $\IT^2$, and are given by
 \beq 
 \Delta_a\ =\ b_a \ \int_{\Gamma} \frac {d^{\,2}\tau }{\tau _2} \,\big(\, Z_{\text{torus}}(\tau,{\bar \tau}\, )-1\, \big) \ ,
 \label{kaplu1}
 \eeq
with 
\beq
Z_{\text{torus}}\ =\ \sum_{m_{1,2},n_{1,2}\in {\bf Z}} e^{\,2\pi i\, \tau \,(m_1n_1+m_2n_2)}
e^{-\pi\tau_2\, M^2}\, ,
\label{kaplu2}
\eeq
and
\beq
M^2 \ =\ \frac {1 }{4\,(\text{Im}\, T)(\text{Im}\,U)}\, |\,n_2TU+n_1T-m_1U+m_2\, |^2 \ .
\label{masilla}
\eeq
Here $n_i,m_i$ are winding numbers and momenta and $M$ is the mass of KK and winding excitations of vector multiplets involved in the loop. We note that $M$ is invariant under $SL(2,{\bf Z})_T\times SL(2,{\bf Z})_U$ by suitable relabeling of the integers $m_{1,2},n_{1,2}$. Performing the integral in (\ref{kaplu1}) one obtains for the 
moduli dependent threshold corrections
\beq
\Delta_a \ =\ -b_a\, \log\Big[\,  (\text{Im}\,T)\, |\eta(T)\, |^4(\text{Im}\,U)|\, \eta(U)|^4\, \Big] \ ,
\eeq
with $b_a$ the $\beta$-function coefficient of the corresponding ${\cal N}=2$ gauge theory.
Note that, as expected, this expression is  invariant under $SL(2,{\bf Z})_T\times SL(2,{\bf Z})_U$.

Let us consider including additional orbifolds to yield 4d $\NN=1$. This leads us to consider $\NN=1$ orbifolds $\IT^6/P$ with the orbifold group $P$ given by $\,Z_N$ and $\IZ_N\times \IZ_M$ orbifolds, with some elements of the orbifold group (forming a subgroup $P_i$) leaving the $i^{th}$ $\IT^2$ fixed. For simplicity we focus on factorizable orbifolds, and consider the dependence on the moduli $T_i$, $U_i$, with $i$ labeling the three possibly fixed $\IT^2$. Using the above results, this is given by
\cite{Dixon:1990pc}
\beq
\Delta_a\ =\ -\sum_i \frac {|P^{(i)}|}  {|P|}\ b_{a,i}^{\, {\cal N}=2}\, \left\{\, \log \left[ \, ( \text{Im}\, T_i)\, |\, \eta(T_i))|^4\, \right]\, +\,\log \left[\, ( \text{Im}\,U_i)\, |\eta(U_i))|^4\, \right] \right\} \ +\ c_a \ ,
\eeq
where $|P|$ is the order of the orbifold group (e.g. $N$ for $\IZ_N$). Also $b_{a,i}^{\,{\cal N}=2}$ is the $\beta$-function coefficient of the corresponding ${\cal N}=2$ sub-theory, and $c_a$ is a moduli-independent constant.  Recall that the orbifold group is constrained to have a crystallographic action on the torus. Also, we note that in certain $\NN=1$ orbifolds, some or all the $U_i$ are frozen and hence only contribute a constant to the thresholds. This occurs when the fixed $\IT^2$ is subject to extra orbifold action which are not $\IZ_2$, and thus act crystallographically only for definite values of the $\IT^2$ complex structure. For instance, that is the case e.g. for the $\IZ_N'$ orbifolds with $N=6,8,12$.

The resulting 4d $\NN=1$ theory contains gauge sectors which may experience strong dynamics effects at low energies. Consider the simplest situation that some gauge factor contains no charged matter, so that it describes a pure super-Yang-Mills sector, which confines and produces a gaugino condensation superpotential in the infrared \footnote{In cases with charged chiral multiplets and possible non-perturbative superpotentials involving them, the modular properties of the superpotential  require including the effect of the modular weights of matter fields \cite{Lust:1990zi,deCarlos:1992kox},
 which in general transform non-trivially under the modular groups. In particular, this also occurs for the tree-level superpotential couplings involving untwisted or twisted fields}. The prototypical case is an unbroken $E_8$, which is present if the embedding of the orbifold group in the gauge degrees of freedom involves only the other $E_8$. This includes standard embedding models, but also many other orbifolds, see \cite{Ibanez:2012zz} for a review.

Including the effect of the threshold correction leads to holomorphic gauge functions
\beqa
f_a \ =\ S\ -\ \frac {1}{16\pi^2}  \ \sum_{i=1}^3\ b_{a,i}^{\, {\cal N}=2}\, \log \left[\, \eta(T_i)^4\, \eta(U_i)^4\, \right] \ ,
\label{kinetic}
\eeqa
where $S$ is the heterotic 4D complex dilaton.
Actually, cancellation of duality anomalies requires in general a Green-Schwarz mechanism 
\cite{Derendinger:1991hq,LopesCardoso:1991ifk,Ibanez:1992hc}, which implies the replacement
$b_{a,i}^{\, {\cal N}=2}\rightarrow b_{a,i}^{\,{\cal N}=2}-\delta_{GS}^i$, where  $\delta_{GS}^ i$ are  gauge group
independent constants. However, this point is not essential for our purposes, so we rather ignore it by considering e.g. the $\IZ_2\times \IZ_2$ orbifold, for which $\delta_{GS}^i=0$.  

The gaugino condensation superpotential then reads
\beq
W_{E_8}\ =\ \Lambda^3  e^{\frac {3f_{E_8}}{2\beta_{E_8}}} \ =\
\Lambda^3 e^{\frac {3S}{2\beta_{E_8}} }\ \prod_{i=1}^3 \frac {1}{\eta(T_i)^2}  \frac {1}{\eta(U_i)^2} \ ,
\eeq
where we have made use of the fact that $\beta_{E_8}=3b_{i,E_8}^{\,{\cal N}=2}$ in this example. 
This has precisely the modular properties described in the previous section, and it is a most simple illustration of how such superpotentials can arise in very explicit heterotic compactifications. 

The result can be easily adapted to other orbifold groups. For instance, in orbifolds where, as mentioned earlier, the complex structure moduli are projected out, they are simply not present in the superpotential. Also, for orbifolds with
$\delta_{\rm GS}^i\not=0$, the only effect is that the exponent of the $\eta(T_i)$'s is smaller  \cite{Lust:1990zi,deCarlos:1992kox}

 Finally, recall that in orbifolds with no fixed planes, like the prime order cases $\IZ_3$, $\IZ_7$, there are no moduli dependent threshold corrections. 
The modular invariant scalar potentials arising from orbifolds with gaugino condensation can be subject to explicit study. For instance, considering examples with fixed complex structure, and focusing on the dependence on the overall K\"ahler modulus, we have a superpotential of the form $W(S,T)= \Omega(S)/\eta(T)^6$, which is one of the simplest examples
whose extrema are studied in section 4 and Appendix B.

As we advanced, these potentials diverge exponentially for large $\text{Im} \ T,\text{Im} \ U$. The origin of this divergence can be traced to contributions from infinite towers of states becoming light, as expected from the swampland distance conjecture. In particular, consider the mass $M^2$ in \ref{masilla} for fixed complex structures: we see that at large $\text{Im} \ T$, the sector with vanishing winding numbers $n_{1,2}=0$ leads to an infinite tower of KK states  with $M^2\simeq ({\rm Im}\,T)^{-1}$. 
As is familiar \cite{OV1,vafafederico,irene,emergence,Lee:2018urn,Lee:2018spm,Hebecker:2018vxz,Grimm:2018cpv,Palti:2017elp},
 the mass decreases exponentially with the distance, if phrased in the proper frame. By $SL(2,\IZ)_T$, the opposite regime of $\text{Im} \ T\rightarrow 0$ leads to an infinite tower of  winding modes with vanishing momenta $m_{1,2}=0$. A similar behavior can be recovered at the points at infinite distance in complex structure moduli space: As $\text{Im} \ U\rightarrow \infty$ (and fixed $T$), there is an infinite tower of states with $m_1=n_2=0$ becoming light. In any of these limits, the physical interpretation is that the towers of light particles modify the gauge theory dynamics by increasing the effective scale of the gaugino condensate, hence leading to a  potential growing at infinity.

The description in terms of the heterotic degrees of freedom certainly resonates with the swampland distance conjecture. In the following section, we consider a dual type I' / Horava-Witten description, which allows for a more precise identification of the degrees of freedom controlling the dynamics of the $T$ and $U$ moduli independently.

\subsection{The type I' or Horava-Witten dual}
\label{sec:hw}

\subsubsection{From heterotic to type I' / HW} 

In the above heterotic description, the piece of moduli space under discussion has an $SL(2,\IZ)_T\times SL(2,\IZ)_U$ modular group. However, the relevant heterotic degrees of freedom correspond to momentum and winding states in $\IT^2$, whose mass formula \ref{masilla}  depends on {\em both} the $T$ and $U$ moduli of $\IT^2$; they are hence not the natural objects to disentangle the two independent $SL(2,\IZ)$ modular properties, or the behavior at independent infinite distance points. In this section we provide a picture in terms of a dual type I' or Horava-Witten description, which explicitly displays the degrees of freedom controlling the dynamics of the $T$ and the $U$ moduli independently.
 
In order to do that, we return to the description of moduli dependent threshold corrections in $\NN=2$ subsectors in terms of a compactification of heterotic on of K3$\times \IT^2$. Again, eventual reduction to $\NN=1$ can be obtained by additional orbifold actions. In this section we would like to focus on the structure of heterotic on K3$\times \IT^2$, and exploit the rich web of string dualities to understand the one-loop threshold correction. 

The compactification can be constructed as a 6d $\NN=1$ compactification on K3, followed by a $\IT^2$ compactification to 4d $\NN=2$.  As is well known, in the $E_8\times E_8$ heterotic, these are determined by the distribution of  instanton numbers for the gauge backgrounds on the two $E_8$ factors,  $(12+n,12-n)$ \cite{Seiberg:1996vs}, while for the $SO(32)$ heterotic, it corresponds to $n=4$ \cite{Morrison:1996na,Berkooz:1996iz}, see \cite{Ibanez:1997td} for a review. We are thus left with a large class of 6d models which should subsequently be compactified on $\IT^2$, possibly with Wilson lines; clearly, this description is disadvantageous since the modular properties in the $\IT^2$ compactification are obscured by the complications of the previous stage of K3 compactification.

It is thus natural to regard the configuration as first a $\IT^2$ compactification down to 8d, eventually followed by a K3 compactification to 4d $\NN=2$. In this description, the modular properties should be manifest already at the level of the toroidal 8d theory. Indeed, the 8d theory contains a precursor of the corrections to the 4d gauge couplings corrections, corresponding to corrections to the 8d couplings $\tr F^4$, $(\tr F^2)^2$ and $(\tr F^2)(\tr R^2)$ (there are in addition $\tr R^4$ and $(\tr R^2)^2$ corrections, not involving the gauge groups, and which we will hence skip). As we show below, the coefficients of such couplings are modular functions of the heterotic K\"ahler modulus $T$ (as well as of its complex structure modulus $U$). Upon dimensional reduction on K3, including the corresponding gauge and gravitational instanton backgrounds, one recovers the 4d $\NN=2$ corrections to the gauge couplings, thus with the precisely (combinations of) the same modular functions.

The structure of the quartic corrections to the 8d theory, for diverse choices of Wilson lines, has been studied from different dual descriptions. For instance, in \cite{Bachas:1997mc,Kiritsis:1997hf,Lerche:1998nx,Lerche:1998gz,Lerche:1998pf,Forger:1998ub,Foerger:1998kw,Gava:1999zk}, either directly as a perturbative computation in heterotic string, or from F-theory on K3. The latter is closely related (in a perturbative limit) to the orientifold of type IIB on $\IT^2$ with 4 O7-planes and 8 D7-branes on top (as counted in the double cover). This corresponds to a heterotic model with Wilson lines breaking the gauge group to $SO(8)^4$, as follows. Starting from the heterotic, one performs an S-duality to a type I $\IT^2$ compactification with Wilson line, and two T-dualities to the type IIB orientifold. 

The original heterotic $T$ modulus maps to the type IIB complexified coupling $\tau$, and the heterotic $U$ modulus remains the complex structure modulus of the type IIB $\IT^2$. 
Hence, the $SL(2,\IZ)^2$ modular group corresponds to S-duality on the type IIB complex coupling $\tau$, and the modular group of the orientifolded $\IT^2$.  In the type IIB picture, the corrections were computed in \cite{Billo:2009di}; here the $\tau$-dependent piece of the quartic corrections arise from D$(-1)$-brane instantons, whereas the $U$-dependent part arises from perturbative fundamental string corrections. Hence, this picture succeeds in disentangling the two modular groups in terms of two kinds of instanton effects. This is very reminiscent of the Swampland Distance Conjecture, since moving towards points at infinite distance in a modulus has an effect in a tower of charge states, albeit in this case they correspond to D$(-1)$-brane instantons rather that light particles.

In order to make the connection with towers of particles as in the standard Swampland Distance Conjecture, we will exploit yet another picture, developed in \cite{Petersson:2010qu} (see also \cite{Gutperle:1999dx}). It corresponds to T-dualizing the type IIB orientifold into an $\IS^1$ compactification of type I' theory\footnote{As will become clear later on, this T-duality is similar to those relating the Weak Gravity Conjecture for charged particles and instantons.}. Namely, we consider type IIA on and $\IS^1/\IZ_2$ orientifold with two O8-planes, each with 16 D8-branes (as counted in the double cover). This 9d theory is then compactified on an additional $\IS^1$ to 8d, e.g. with suitable Wilson lines if we wish to relate to the $SO(8)^4$ theory. In fact, the discussion can be easily extended to general Wilson lines, as we will do later on.

In this type I' picture, the threshold correction is associated to the tower of particles in the 9d type I' theory, concretely the corrections depending on $\tau$ arise from D0-brane particles bound to each of the two O8/D8 configurations. As discussed in \cite{Kachru:1996nd} the spectrum of such BPS particles is given by bound states of $2k$ D0-branes, giving particles in the ${\bf 120}$ of each $SO(16)$, and bound states of $(2k+1)$ D0-branes, giving particles in the ${\bf 128}$ spinor representation of $SO(16)$. The 8d correction arising as a 1-loop diagram of these particles, in which they are allowed to run along the $\IS^1$ bringing us down from 9d to 8d. Similarly, the corrections depending on $U$ arise from 9d particles arising from open string stretching among the D8-branes, and thus having non-trivial winding in the type I' $\IS^1\IZ_2$. These are 9d particles with winding $w\in\IZ$ and transforming in the $({\bf 16},{\bf 16})$ of $SO(16)^2$, or with winding $2w$ and transforming in the $({\bf 120},{\bf 1})+({\bf 1},{\bf 120})$.

In this picture, the modular behaviour in $\tau$ is not manifest. Indeed, since $\tau$ is the type IIB complex coupling, the corresponding S-duality group is manifest only if we lift up to M-theory. The lift of type I' theory corresponds to Horava-Witten (HW) theory, in the particular choice of wilson lines breaking the $E_8\times E_8$ symmetry on the boundaries down to $SO(16)^2$. In this picture the D0-branes lift to KK momentum modes of the 10d $E_8$ gauge multiplets, which propagate on a $\IT^2$, given by one circle associated to the lift from type I' to 11d, and the second circle being that already present in the $9d\to 8d$ compactification of type I'. As in standard M-theory/type IIB duality, the complex structure $\tau$ of this $\IT^2$ maps to the type IIB complex coupling, and there is a geometric interpretation for the $SL(2,\IZ)$.
Similarly, the $U$-depended contribution in type IIB theory arises from M2-branes wrapped on the $\IS^1/\IZ_2\times \IT^2$ geometry.

The dual pictures and relevant ingredients are shown in Table \ref{latabla}.

\begin{table}
\begin{center}
	\begin{tabular}{|c|c|c|c|}
		\hline 
		Picture	& Modulus &  Duality group & Objects \\
		\hline 
		\hline 
		Heterotic & $\int_{\IT^2} B_2+i\, R_9 R_{10}$ &  K\"ahler  T-duality & Mix of \\
		\cline{2-3}
		& $R_{10}/R_9 e^{i\theta}$ & Complex structure & wrapped F1 / KK \\
		\hline\hline
		Type I & $\int_{\IT^2} C_2+i\, R_9 R_{10}/g_s$ &  K\"ahler  T-duality & Wrapped D1\\
		\cline{2-4}
		& $R_{10}/R_9 e^{i\theta}$ & Complex structure  & KK \\
		\hline\hline
		Type IIB & $C_0+i\, 1/g_s$ & S-duality & D$(-1)$\\
		\cline{2-4}
		orientifold & $R_9/R_{10} \,e^{i\theta}$ & Complex structure  & Wrapped F1  \\
		\hline\hline
		Type I' & $\int_{\IS^1}C_1+i\, R_9/g_s$ &  Hidden HW & D0\\
		\cline{2-4}
		orientifold & $\int_{\IS^1\times I} B_2+i\, R_9 R_{10}$ & K\"ahler  T-duality & F1 winding \\
		\hline\hline
		Horava-Witten & $R_{11}/R_9\, e^{i\theta}$ & Complex structure & KK\\
		\cline{2-4}
		orientifold & $\int_{\IT^2\times I} C_3 + i R_9R_{10}R_{11}$ & K\"ahler  T-duality & wrapped M2  \\
		\hline\hline
	\end{tabular}\caption{Dual pictures and some of their ingredients. Despite similar notation for the compactification radii, they in general have meaning adapted to the corresponding picture. }
	\label{latabla}
\end{center}
\end{table}

\medskip

\subsubsection{The 1-loop computation of the threshold correction}

Thus, HW theory on $\IT^2$, possibly with Wilson lines, is the natural framework to understand the modular properties of the threshold corrections. To flesh out the discussion, in the following we illustrate the computation of the $\tau$-dependent threshold correction for the configuration with an unbroken $E_8$ in 8d. Cases with more general Wilson lines can be discussed similarly. We follow closely the steps in \cite{Petersson:2010qu}, albeit for the $E_8$ case.

The 1-loop computation of the $E_8$ HW vector multiplets, described from the perspective of the worldline, reads  
\begin{eqnarray}
\label{Mgauge}
\mathcal{A}^{gauge}  &=& \frac{1}{4!}
\int_{0}^{\infty} \frac{dt}{t} t^4 \sum_{\ell_I} \int d^{8}\mathbf{p} ~ e^{-\pi t \left( \mathbf{p}^2+G^{IJ} \widetilde{\ell}_{I} \widetilde{\ell}_{J}\right)} \nonumber\\ 
&=&\frac{1}{4!}
 \int_{0}^{\infty} \frac{dt}{t} \sum_{\ell_9 , \ell_{11}}~ e^{-\pi t  \frac{1}{V_{(2) }\tau_2} | \widetilde{\ell}_{9} -\tau\widetilde{\ell}_{11} |^2}
\end{eqnarray}
Here ${\widetilde \ell}_I= \ell_I - {\bf \Lambda}\cdot {\bf A}_I$, with $\Lambda$ a root vector of $E_8$ and $ {\bf A}_I$ the Wilson line in the $I^{th}$ direction. We focus on the case without Wilson lines, and drop the tildes for the momenta such that $\widetilde{\ell}_9=\ell_9\in\IZ$, $\widetilde{\ell}_{11}=\ell_{11}\in\IZ$. 
The above expression for vanishing Wilson lines is manifestly modular invariant under transformations of $\IT^2$. It is also straightforward to analyze the surviving subgroups in the presence of non-trivial Wilson lines, like the $SO(8)^4$ or $SO(16)^2$ points. Moreover, for general Wilson lines, there are combined identifications due to modular transformations of $\IT^2$ accompanied by non-trivial actions on the Wilson line moduli. This is just the HW description of the dualities in the heterotic $O(2,18;\IZ)\backslash O(2,18;\IR)/[O(2;\IR)\times O(18,\IR)]$ 

In the above expression, the $t^4$ in the integral is associated to the four external legs in the $\tr F^4$ term. Note that the same result is obtained by taking two external legs i.e. $\tr F^2$ in the 4d theory.

In the following we reabsorb $V_{(2)}$ in a rescaling of $t$. A useful way to eventually extract the dominant contribution at infinity in $\tau$ is to perform a Poisson resummation to get
\begin{eqnarray}
\label{Mgauge2}
\mathcal{A}^{gauge} &=& \frac{\sqrt{\pi}}{4!} 
\int_{0}^{\infty} \frac{dt}{t} t^{-1/2} \sum_{w_9 , \ell_{11}\in\IZ} 
e^{- \frac{\pi^2  w_{9}^{2} }{t} -  \tau_{2}^{2} {\ell}_{11}^{2}  t   }  ~
e^{2\pi i w_9 {\ell}_{11} \tau_1}  ~.
\end{eqnarray}
In the 10d picture upon reduction on the direction 11,  the integer $w_9$ represents the winding of the 9d objects (eventually D0-branes) running in the loop in the $\IS^1$ bringing us down from 9d to 8d. 

Isolating the $w_9=0$ contribution, and performing a Poisson resummation in $\ell_{11}$ and integrating in $t$, we have
\begin{eqnarray}
\label{treegauge}
\mathcal{A}^{gauge}_{(w_{9}=0)}
&=& \frac{\tau_2}{4!\pi} 
 \sum_{ w_{11}\neq 0} \frac{1}{w_{11}^{2}} = \frac{\pi\tau_2}{72}
\end{eqnarray}
where we have exclude the divergent $w_{11}=0$ term (discussed below).

As will be clear later on, this contribution extracts the dominant contribution in the large $\tau$ limit of an infinite tower labeled by  $\ell_{11}$, of D0-branes in type I' language. With hindsight we write
\begin{eqnarray}
\label{treegauge3}
\mathcal{A}^{gauge}_{(w_{9}=0)} = -\frac1{12}\left(\, \frac{2\pi i \tau}{24} -  \frac{2\pi i {\ov\tau}}{24} \right)=-\frac 1{12} (\log q^{\frac{1}{24}} + \log {\ov q}^{\frac{1}{24}})
\end{eqnarray}
Incidentally, this zero-winding contribution from 11d KK modes is similar to a contribution from D0-branes in the Gopalumar-Vafa interpretation of the topological string \cite{Gopakumar:1998jq}.

The $w_9\neq 0$ contribution, after integration in $t$ reads
\begin{eqnarray}
\label{gaugecorr}
\mathcal{A}^{gauge}_{(w_{9}\neq 0)}
=\frac{1}{4!}  
\sum_{{}^{w_9 \neq 0}_{\ell_{11} \in {\mathbf Z}}} \frac{1}{|w_9 | } 
e^{-2\pi \tau_2 |w_9 \ell_{11}|} ~
e^{2\pi i \tau_1 w_9 \ell_{11}} ~.
\end{eqnarray}
Introducing $q=e^{2\pi i \tau}$, and gathering the contributions from D0's ($\ell_{11}>0$) and anti-D0-branes ($\ell_{11}<0$), we have
\begin{eqnarray}
\label{gaugecorr}
\mathcal{A}^{gauge}_{(w_{9}\neq 0)}
=\frac{2}{4!}  \Bigg(\, \sum_{{}^{w_9> 0}_{\ell_{11}>0}} \frac{1}{w_9  } q^{w_9\ell_{11}} \,+\, cc.\, \Bigg)
\end{eqnarray}
Here have substracted the $\ell_{11}=0$ piece, which corresponds to no D0-branes, and corresponds to a perturbative piece mentioned later on. The above expression can be recast as
\beqa
\mathcal{A}^{gauge}_{(w_{9}\neq 0)}
= -\frac 1{12} \sum_{\ell_{11}>0} \log (1-q^{\ell_{11}})\, +\, cc. \, =  -\frac 1{12} \log \left[ \,\prod_{\ell_{11}>0} (1-q^{\ell_{11}}) \right]\, +\, cc.
\eeqa
So putting together with (\ref{treegauge3}) we have
\beqa
\mathcal{A}^{gauge}=-\frac 1{6} \log |\eta(\tau)| = -\frac 1{24}  \log |\eta(\tau)|^4
\eeqa
Here the exponent is chosen by adjusting the coefficient, but it can be checked that it matches with the $\log ({\rm Im}\, \tau)$ piece (arising from the  dropped $\ell_{11}=0$ part) to achieve a modular invariant quantity, as expected from the original manifestly modular invariant amplitude in the HW picture. Hence we get
\beqa
\log \, [\, ({\rm Im}\, \tau) |\eta(\tau)|^4\,]
\eeqa

As mentioned before, the dominant term at large $\tau$ corresponds to the zero winding $w_9=0$ contribution of the whole tower of D0-branes (or HW KK modes in the 11 direction). It is thus these states that dominate in the point at infinity in the moduli space of $\tau$ (namely, $T$ in the heterotic description).

A very similar computation involving fundamental string winding states in type I' theory (i.e. wrapped M2-brane states in HW) can be shown to produce the $U$-dependent threshold corrections. These were discussed in \cite{Petersson:2010qu} in the type I' picture for the $SO(16)^2$ and $SO(8)^4$ theories, to which we refer the reader for details. Here it suffices to note that, from the perspective of the 9d theory, the computation is identical to the previous one, as it simply relies the existence of one 9d BPS state for each integer charge $k\in \IZ$; in the present case, the relevant 9d states are M2-branes with wrapping number $k$ on the two-dimensional space defined by the HW interval times the $\IS^1$ in the lift type I'$\to$ HW. This multiplicity of bound states implies the result
\beqa
\log \, [\,(\,{\rm Im}\, U)\, |\eta(U)|^4\,]
\eeqa

\medskip

Let us conclude with a final remark bringing us back to the original discussion of heterotic in K3$\times \IT^2$.
The above computation produces the coefficient of the $(\tr F^2)^2$ term in 8d (recall that $E_8$ has no quartic Casimir). One can similarly get the $(\tr F^2)(\tr R^2)$ term, resulting in a quartic correction of the form
\beqa
\log\, [\, ({\rm Im}\,\tau) |\eta(\tau)|^4\, ({\rm Im} \,U\,)\,|\eta(U)|^4\, ]\, (\, \tr F^2\, -\, \tr R^2\,)\, \tr F^2
\eeqa
Upon compactification on K3, with a gauge instanton background with instanton number $(12+n)$, and taking into account the Euler characteristic of K3, we have a threshold correction to the 4d gauge coupling constants given by
\beqa
\log\, [\, ({\rm Im}\,\tau) |\eta(\tau)|^4\, ({\rm Im} \,U\,)\,|\eta(U)|^4\, ]\, (n-12)\, \tr F^2
\eeqa
which, as announced at the beginning, is controlled by the beta function of the 4d $\NN=2$, equivalently the coefficient of the anomaly polynomial of the 6d $\NN=1$ theory obtained from just the K3 compactification.

An equivalent way to understand this picture is to indeed consider first the compactification on K3 to 6d. In this compactification, the tower of D0-branes in the adjoint of $E_8$ will produce a  tower of 6d (massive) D0-brane states, in diverse representations of the surviving 6d gauge group. The ground states in the dimensional reduction in K3 are determined by the index theorem on K3 for states coupled to the gravitational and gauge backgrounds. These massive 6d states are subsequently compactified on $\IT^2$, and produce, via a one-loop diagram whose computation is identical to the above one, the threshold corrections to the 4d gauge couplings. The matching with the above description is precisely due to the familiar fact that the index theorem relates the multiplicities of 6d states to the integral of $ \tr F^2\, -\, \tr R^2$.

\subsubsection{The gaugino condensate}

Let us conclude by mentioning that the type I' or HW picture also includes the source of the non-perturbative effects, which in the 4d $\NN=1$ context produces the non-perturbative superpotential. The relevant object is a 6d BPS particle given by a D4-brane wrapped on K3, and which runs in a loop in $\IT^2$. In the 4d $\NN=2$ context this gives rise to a tower of spacetime instantons, with (in their Higgs branch) correspond to gauge instantons on the D8-branes. In the 4d $\NN=1$ setup, these instanton would have too many fermion zero modes to contribute to the superpotential, and only a few (fractional) instantons provide non-trivial contribution, thanks to the extra orbifold projections. Thus the 1-loop diagram picture is not so useful. In any event, it is interesting to point out that this description shows that the effect of the threshold corrections on the non-perturbative superpotential is nothing but the sum of the contributions from polyinstantons like those introduced in \cite{Blumenhagen:2012kz}, namely the effect of instantons from D0-brane loops on the action of an instanton from a D4-brane loop. 

In the HW lift, we have just an instanton effect from an M5-brane on K3 times the 11d circle, with arbitrary momentum excitations on the latter. This configuration is similar to that in \cite{Bachas:1999um}, but with the additional difficulty that the M5-brane is bound to the $E_8$ boundary, which thus renders the quantitative description beyond present knowledge.

\section{Modular invariant potentials and the swampland conjectures}

\subsection{Extrema in a Modular Invariant Potential}
\label{sec:potential}

In the previous sections we have seen how non-perturbative effects are able to generate $\NN=1$ superpotentials which are modular forms under $SL(2,{\bf Z})$, keeping invariant the moduli scalar potential. In this section we explore in some detail the extrema of the most general class of modular invariant potentials for a single modulus $T$. These of course are simplified models, since a fully realistic string compactification is expected to involve multiple moduli, gauge bosons and matter fields. So at best, these models could arise in some particular compactification after the rest of the spectrum has been integrated out, or could perhaps represent some subsector of a more complicated compactification. On the other hand one could ignore its possible string theory origin and consider them by themselves as explicit examples of $\NN=1$ supergravity models with one chiral field and admitting 
a consistent coupling to quantum gravity. The idea here is to test in specific models the connection between duality symmetries in string compactifications  and the swampland ideas. Thus e.g. one may ask whether the properties of modular invariance imply or are related to any of the proposed swampland constraints.

Let us consider then  the $\NN=1$ supergravity K\"ahler potential 
\beq 
G(T,T^*)\ =    -\,3 \log\,(T-T^*) \,+\, \log|W|^2 \ ,
\label{potone}
\eeq
with a general superpotential $W=H(T)/\eta(T)^6$, with $H(T)$ given by eq.(\ref{generalH}). The corresponding potential is given by
\begin{equation}
V\left(T,T^{*}\right)=\frac{ 1}{8 T_{I}^{3}\left|\eta\right|^{12}} \left\{ \frac{4T_{I}^{2}}{3}\left|   \frac{dH}{dT}+\frac{3}{2\pi}H\hat{G}_{2}\right|^{2} \
-3\left|H\right|^{2} \right\}  \ .
\end{equation}
Here we have renamed  the real and imaginary parts of the field as: $T=\theta+it=T_{R}+i T_{I}$. The potential is invariant under $PSL(2,{\bf Z})$ with the fundamental region in  the complex plane  shown in  fig.(1).  One can show then that the points $T=i$ and $T=\rho=e^{i\frac{2\pi}{3}}$ are always extrema, since they are fixed points under order finite order subgroups which act non-trivially on derivatives of the potential, which must thus vanish.  In \cite{Cvetic:1991qm} it was also conjectured  that all extrema lie on the boundary of the fundamental region.  We have explored the potential numerically and we find that this is indeed the case, although we are not aware of a general proof.  
  
We have performed a general study of minima for the different models obtained for different $m,n\geq 0$ and polynomial ${\cal P}$ in eq.(\ref{generalH}). The summary of our results is that we find Minkowski and AdS extrema with and without SUSY. We also find dS maxima and saddle points, but we have been unable to find dS minima. It is natural to believe that this structure is a consequence of the modular invariance (duality in a string setting) of the models.
  
Although we do not have a general proof for the absence of dS minima, one can  explicitly prove that the extrema at the self-dual points or zeros of ${\cal P}$ are never dS minima. This may be proved for any value of $n,m$ and choice of polynomial ${\cal P}(j(T))$. The proof borrows some results from \cite{Cvetic:1991qm}. First we study, for each choice of $H$, which type of extrema is generated at these points. Once we have
   found a set of parameters generating  a minimum, we show that, for these  parameters, it can never be in dS.  Let us see how his comes about, considering different values of
  $n,m$ in turn.

\begin{figure}[H]
	\centering{}
	\includegraphics[scale=0.39]{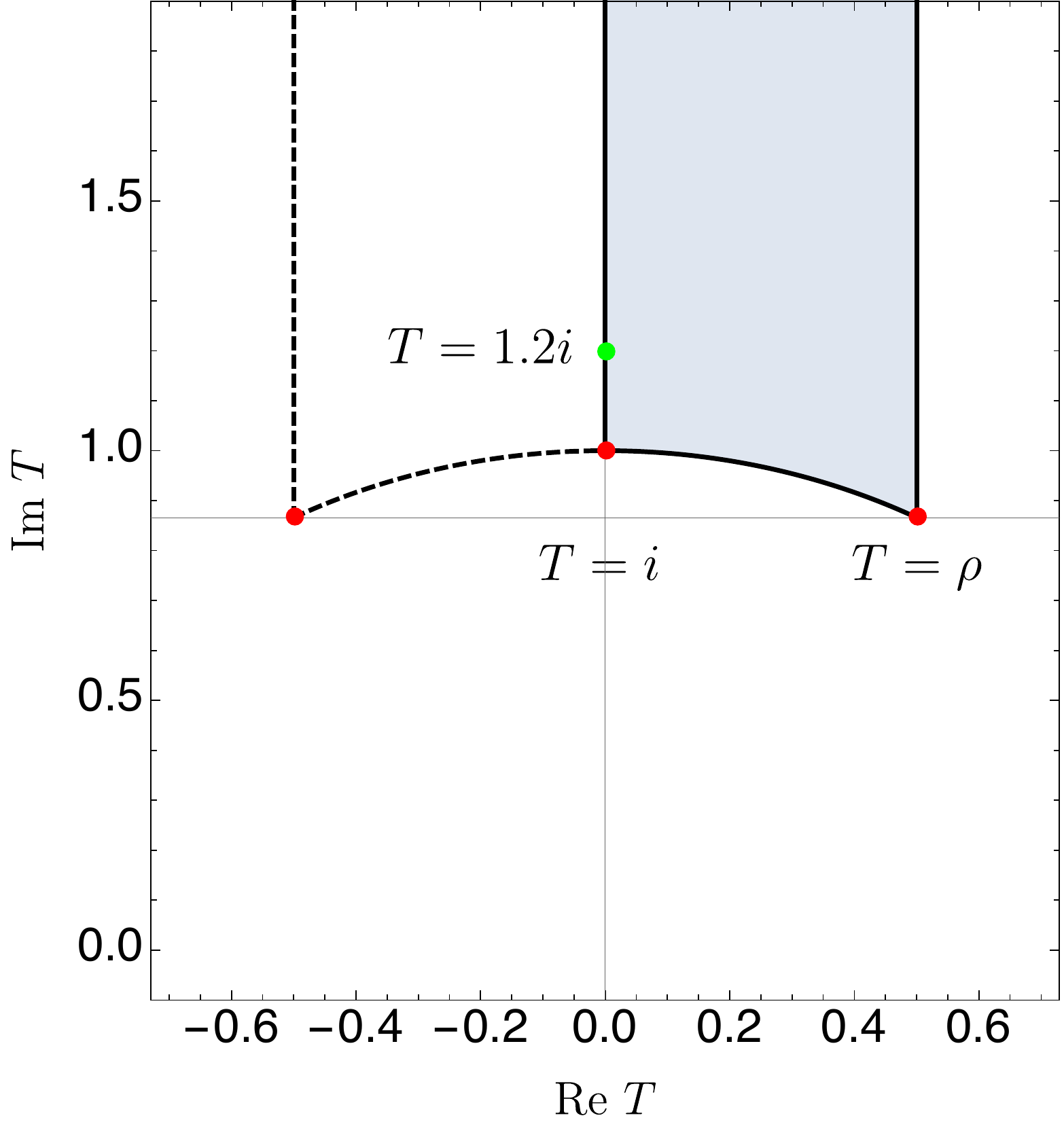}
	\label{domain}
	\caption{\footnotesize Fundamental domain for $PSL(2,{\bf Z})$. The self-dual points are located at  $T=i,\rho$. Other extrema are found on the border of the modular domain.}
\end{figure}

\begin{itemize}

\item {$n=0$ or $m=0$}

At $T=\rho$ one has $\hat{G}_{2}=\frac{dj}{dT}=0$ so that for  $n=0$ one has  $\frac{dH}{dT}=0$ and
\begin{equation}
V\left(T,T^{*}\right) = \frac{\left|  {\cal P}\left( 0 \right)  \right|^{2}}{8  T_{I}^{3}\left|\eta\right|^{12}}\left\{-3\right\}.
\end{equation}
It is always a maximum and it is always in AdS.

At $T=i$ one has  $j=1728$, $\hat{G}_{2}=\frac{dj}{dT}=0$, $ \left| H \right|=1728^{m/2} \left| {\cal P}\left( 1728 \right)\right| $ so for $m=0$ $\frac{dH}{dT}=0$ and one gets 
\begin{equation}
V\left(T,T^{*}\right) = \frac{ \left| {\cal P}\left( 1728 \right) \right|^2}{8  T_{I}^{3}\left|\eta\right|^{12}}\left\{-3\right\} 
\end{equation}
By choosing different $\cal P$ all types of extrema can occur at $T=i$: maximum, minimum or saddle point. Namely,  it is a maximum  if $-2.57<\frac{H'''}{H'}<-1.57$, a saddle point if $\frac{H'''}{H'}<-2.57$ or $\frac{H'''}{H'}>-1.57$ and a minimum if $\left|\frac{H''}{H}+1.19\right|>\frac{3}{2}$. Again it is always in AdS.
In Fig. \ref{cerocero} we consider the particular case $n=m=0$ with  ${\cal P}=1$. We can see the maximum at $T=\rho$ and a saddle point at $T=i$. Additionally, there is a 
non-SUSY AdS minimum at $\text{Im} \ T=1.2$, in the boundary of the fundamental region, but not at a fixed point.
  This is the simplest example superpotential discussed above with $W=1/\eta(T)^6$.

\begin{figure}[H]
	\centering{}
	\includegraphics[scale=0.39]{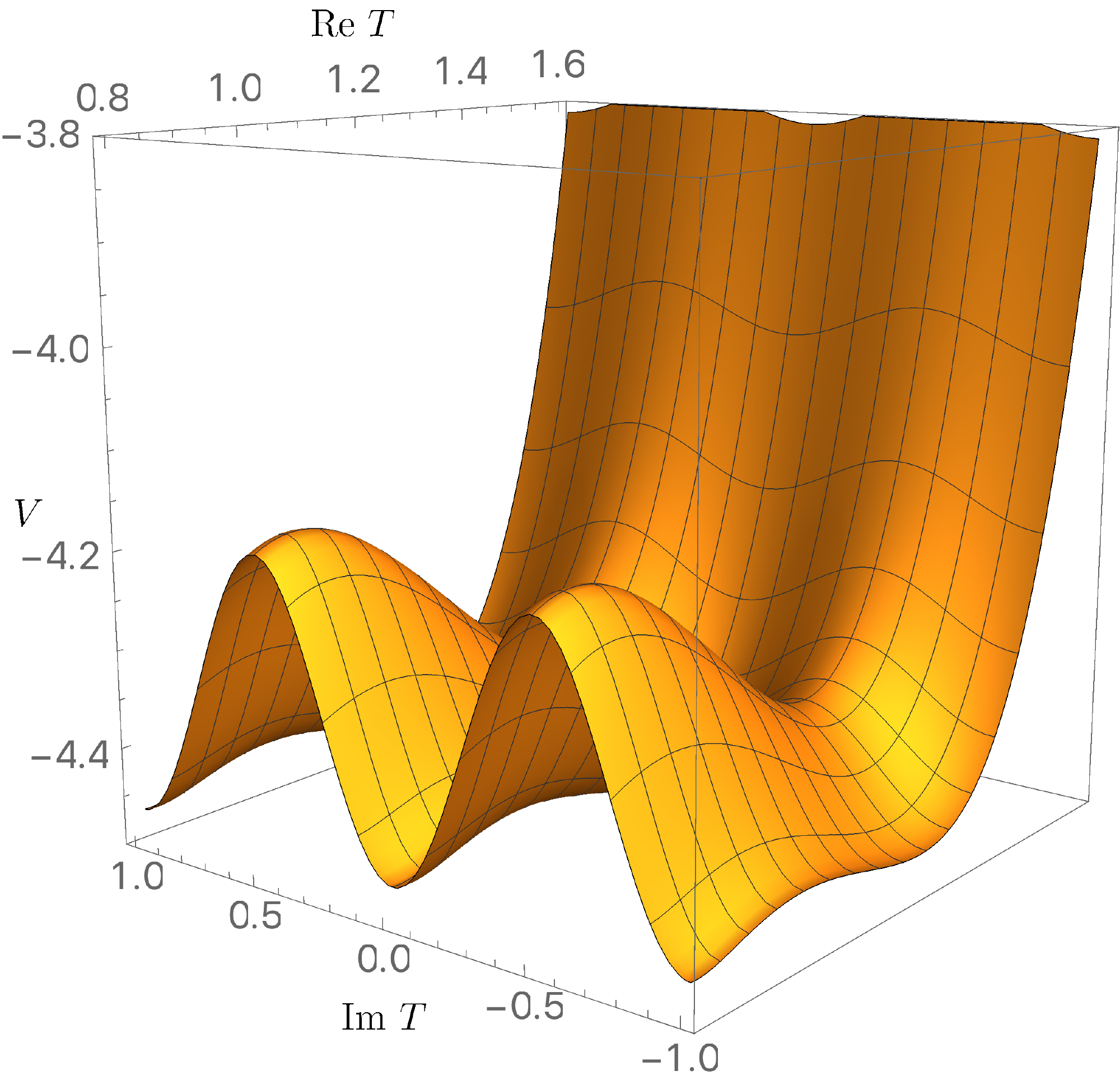}
	\caption{\footnotesize Scalar potential for $W=1/\eta(T)^6$ (i.e. $n=m=0$, ${\cal P}=1$). There  is a SUSY  AdS  maximum at
	$T=\rho$, a saddle point at $T=i$ and a non-SUSY AdS minimum at $\text{Im} \ T=1.2$, which is also on the boundary of the fundamental domain.  }
	\label{cerocero}
\end{figure}


\item {$n>1$ or $m>1$}

These always yield Minkowski minima. If $n>1$  then $T=\rho$ is always a minimum, and the same applies for $m>1$ and $T=i$.   If $n>1$ then $H\left( \rho \right)=0$ and if $m>1$ then $H\left( 1 \right)=0$. In both points $\hat{G}_{2}=\frac{dj}{dT}=0$. If $n>1$ then at $T=\rho$ $\frac{\partial H}{\partial j}$ does not diverge, so $V=0$. Thus 
indeed it is always a Minkowski minimum, and the same happens for $m>1$ at $T=i$.

\item {$n=1$ or $m=1$}

Some of these give rise to dS maxima, but no minima.
If $n=1$ then, at $T=\rho$: $H=\hat{G}_{2}=0$. However, $\frac{\partial H}{\partial j}$ diverges in such a way that $\frac{d H}{dT}$ is finite and non-zero. 
Then the potential, at both  fixed points, simplifies to:
\begin{equation}
V= \frac{1}{8 T_{I}^{3}\left|\eta\right|^{12}}\left\{ \frac{4T_{I}^{2}}{3}\left|\left(\frac{n}{3}(j-1728)^{m/2}j^{-1+\frac{n}{3}}+\frac{m}{2}(j-1728)^{-1+\frac{m}{2}}j^{n/3}\right){\cal P}\left(j\right) \frac{dj}{dT}\right|^{2}\right\} 
\end{equation}

For $n=1$, at $T=\rho$ one obtains
\begin{equation}
V = \frac{2 \left| {\cal P} \left( 0 \right)  \right|^{2}}{3 T_{I}\left|\eta\right|^{12}} \left\{\left|C\right|^{2}\right\}  \ ,
\end{equation}
where we have defined $C=\lim_{T\rightarrow\rho} \frac{1}{3}j^{-\frac{2}{3}}\frac{dj}{dT}$.
This  is always a maxima and the potential is positive. 
In Fig \ref{mceronuno} we plot the potential for $n=1$, $m=0$, ${ \cal P} = 1 $. We can see the predicted dS maximum at $T=\rho$ in the zoom in around this point.  In the same plot, at $T=i$ there is an AdS minimum, since $m=0$. 

Moving on to $m=1$, we evaluate the potential at $T=i$:
\begin{equation}
V = \frac{2 \left| {\cal P} \left( 1728 \right)  \right|^{2}}{3 T_{I}\left|\eta\right|^{12}} \left\{\left|D\right|^{2}\right\}  \ ,
\end{equation}
where we have defined $D=\lim_{T\rightarrow i}$$\frac{1}{2}\left(j-1728 \right)^{-\frac{\text{1}}{2}}\frac{dj}{dT}$.
The potential is always positive at the extrema. By changing H we can make it a saddle point or a maximum but not a minimum.

\end{itemize}

The remaining extrema are the multiple zeros of ${\cal P}\left(j\right)$. They verify $H= \frac{dH}{dT}= 0$,
 so they preserve supersymmetry and  are always Minkowski.
We summarize the results for minima at the self-dual points in the tables (\ref{tablaone}) and (\ref{tablatwo}).
Note that in the model considered supersymmetry is spontaneously broken if at the minimum of $V$ the auxiliary field
\begin{equation}
h^{T}\propto\ \left|H\right|\left(\frac{1}{H}\frac{dH}{dT}+\frac{3}{2\pi}\hat{G}_{2}\right)\rightarrow\  \left(\frac{dH}{dT}\right)
\end{equation}
has a non-zero vacuum expectation value. In the last step we specified to the self-dual points, where $\hat{G}_{2}=0$.

This completes the study of extrema on self-dual points, which are generic.

\begin{figure}[H]
	\centering{}
	\label{mceronuno}
	\includegraphics[scale=0.35]{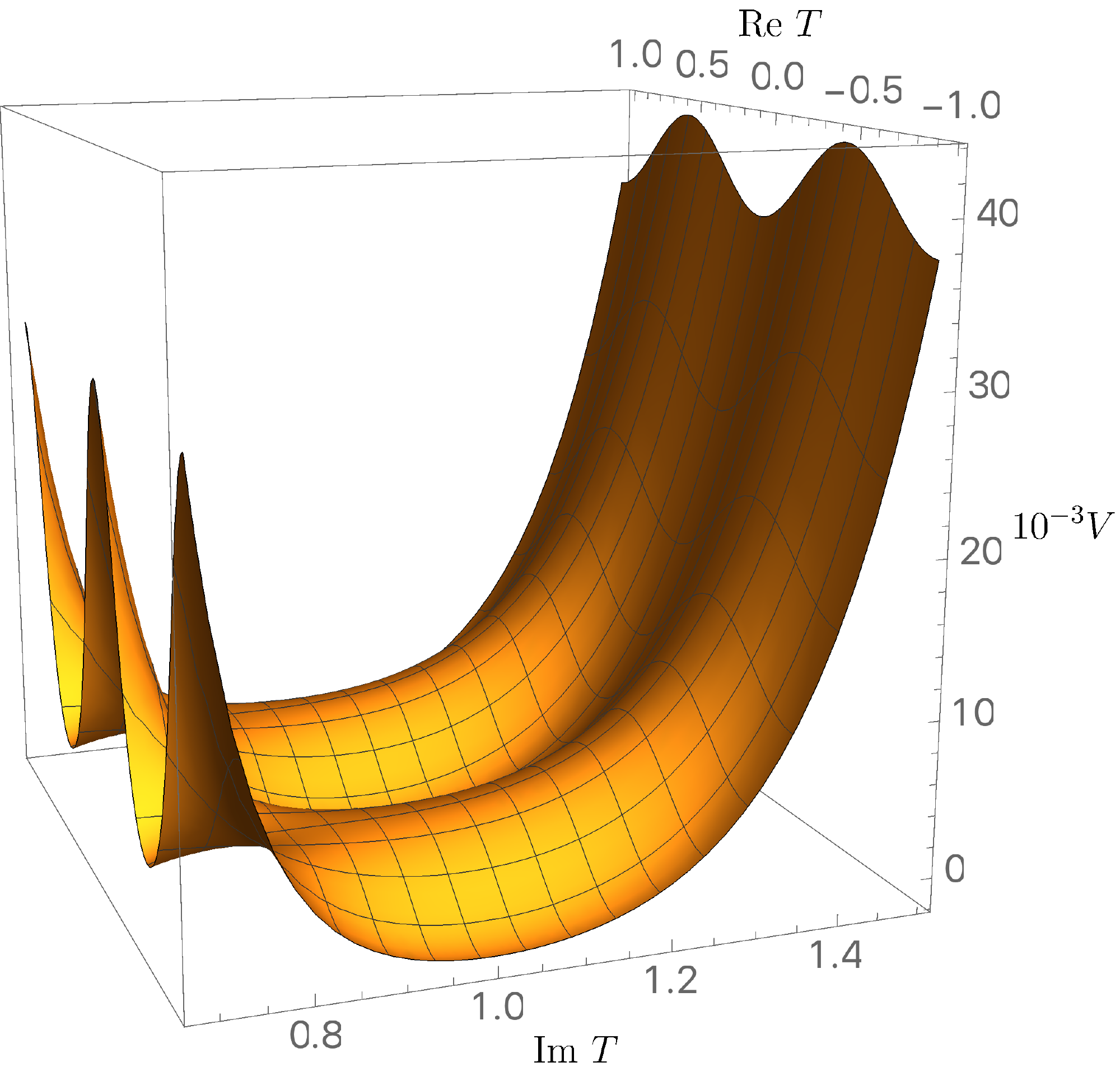} 
	\includegraphics[scale=0.35]{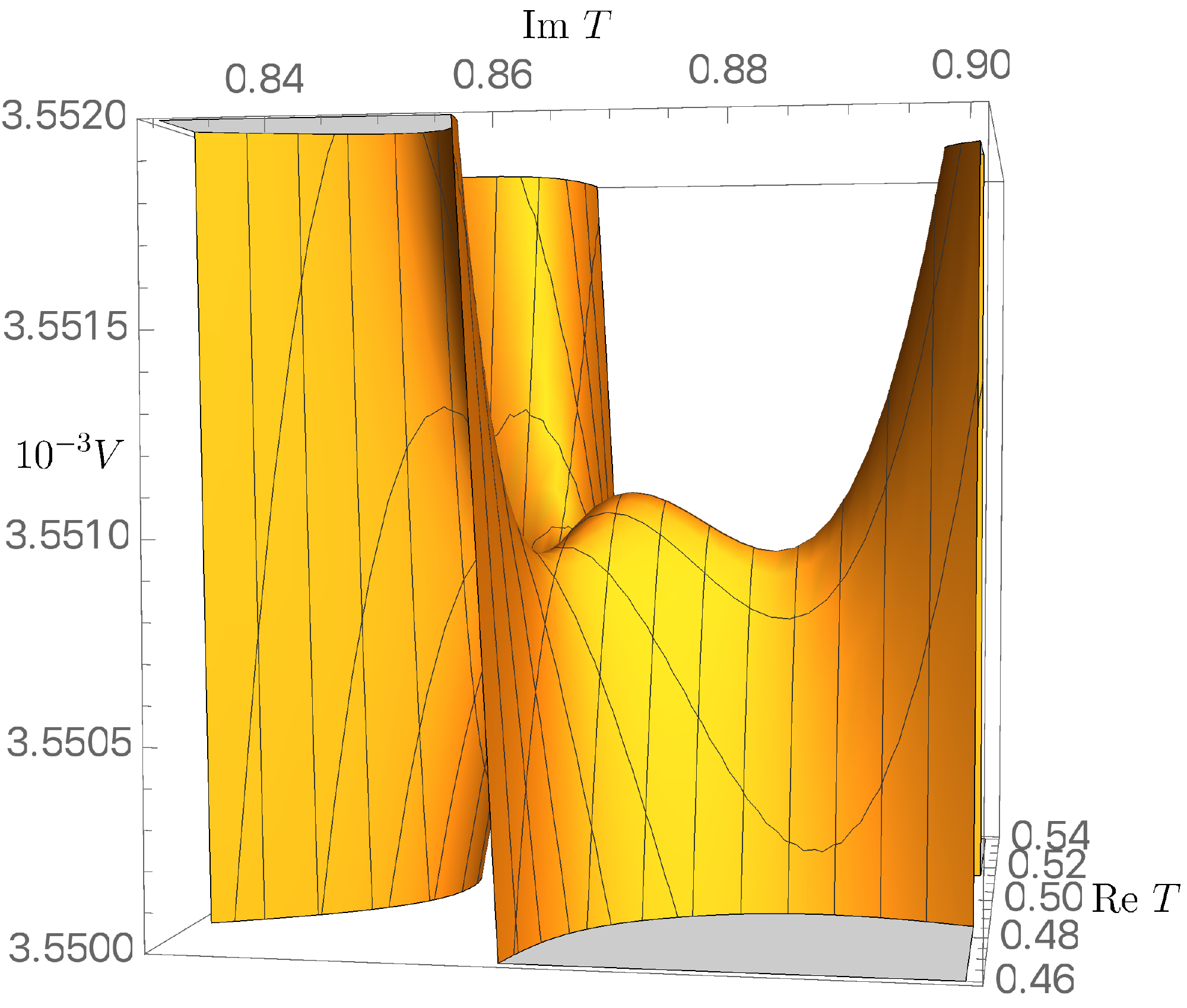}
	\label{one}
	\caption{\footnotesize Left: Scalar potential for $W=j^{1/3}/\eta^6$  (i.e. $n=1$, $m=0$, ${\cal P} =1$): Right:  A zoom around its dS maximum.}
\end{figure}

\begin{table}[H]
	\begin{tabular}{|c|c|c|c|c|c|}
		\hline 
		& $V\left(T=1\right)$ & Type of Extrema & $H$  &$ \frac{dH}{dT}$ & SUSY\tabularnewline
		\hline 
		\hline 
		$m>1$ & $V=0$ & Min  & $0$ & $0$ & Yes\tabularnewline
		\hline 
		$m=1$ & $ \frac{1}{T_{I}^{3}\left|\eta\right|^{12}}\left\{ \left|a\right|^{2}\left|C\right|^{2}\right\} >0$ & Max $-2.57<\frac{H'''}{H'}<-1.57$ & $0$ & $\neq 0$ & No \tabularnewline
		&  & SP $\frac{H'''}{H'}<-2.57$ or $\frac{H'''}{H'}>-1.57$ & & & \tabularnewline
		\hline 
		$m=0$ & $\propto\frac{\left|P(0)\right|^{2}}{T_{I}^{3}\left|\eta\right|^{12}}\left\{ -3 \right\} < 0 $ & Min $\left|\frac{H''}{H}+1.19\right|>\frac{3}{2}$ & $\neq0$ & $0$& Yes \tabularnewline
		&  & Max $-\frac{3}{4}<\frac{H''}{H}+1.19<\frac{3}{4}$ & && \tabularnewline
		&  & SP (Saddle Point) if else &&& \tabularnewline
		\hline 
	\end{tabular}\caption{Classification of the extrema found at $T=i$.}
	\label{tablaone}
\end{table}
\begin{table}[H]
	\begin{tabular}{|c|c|c|c|c|c|}
		\hline 
		& $V\left(T=\rho\right)$ & Type of Extrema & $H $ &$ \frac{dH}{dT}$ & SUSY\tabularnewline
		\hline 
		\hline 
		$n>1$ & $V=0$ & Minimum  & $0$ & $0$ & Yes \tabularnewline
		\hline 
		$n=1$ & $ \frac{1}{\left|\eta\right|^{12}}\left\{ \frac{4}{3}\left|{\cal P}\left(1728 \right)\right|^{2}\left|D\right|^{2}\right\} >0$ & Maximum  & $0$ & $\neq 0$ & No \tabularnewline
		\hline 
		$n=0$ & $\propto\frac{1728^{m}\left|{\cal P}\left(1728\right)\right|^{2}}{T_{I}^{3}\left|\eta\right|^{12}}\left\{ -3 \right\} < 0 $ & Maximum & $\neq0$ & $0$ & Yes\tabularnewline
		\hline 
	\end{tabular}\caption{ Classification of the  extrema found at $T=\rho$.}
	\label{tablatwo}
\end{table}

It is quite interesting  that modular functions conspire to produce all kinds  of extrema except dS minima. 
One may wonder why, given the freedom in $H$ and ${\cal P}$ we are  still  unable to uplift one of the Minkowski vacua that we have found to dS. 
One would naively say that, starting from a Minkowski vacuum in a self-dual point and adding some perturbation there should be some nearby dS minima, not only AdS.
This is in fact not the case, and adding (modular invariant) perturbations one never lands in dS.
The whole potential is constructed around the Klein Invariant function $j(T)$ and the Dedekind function $\eta(T)$, as a consequence of modular invariance. 
The dependence on $\eta$ is fixed and we have only some freedom in choosing the dependence on $j$. It is easy to see why using only this freedom the Minkowski or AdS minimum cannot be ``perturbed" to a dS minima. If $j$ or $(j-1728)$ are zero at the minimum nothing changes by adding them as perturbations. We believe that something similar happens at the other points in the boundary of the fundamental modular region, but we have been unable to prove it  analytically. 
Numerically we have checked that the other minima in the boundary, which appear for some specific choices of $H$ are never dS minima.

\subsection{Swampland conjectures}

We have already seen that modular invariant potentials have features consistent with the swampland ideas and how such potentials
appear in models of reduced supersymmetry at the non-perturbative level. In particular we have seen how for large values of the
moduli, modular invariance dictates the existence of a dynamical barrier censoring large moduli which would allow for global symmetries.
The origin of such a barrier in the specific string models studied correspond to towers of states becoming massless, in agreement with 
the distance conjecture ideas. The towers of states are KK, winding, $D0$  or instanton modes depending on the specific model.
We have found dS maxima, in violation of the original dS swampland conjecture of ref.\cite{Obied:2018sgi}. However, in all examples analyzed the 
refined conjecture \cite{Ooguri:2018wrx}  including the third condition in the beginning of this paper,  seems to be obeyed.
 We have checked it numerically for each case by plotting
 \beq
  \frac {\left| \nabla V\right|} {V} \ =\ \frac {\sqrt{K^{i\overline{j}}\partial_{i}V\partial_j{\overline V}   }}       {V} 
  \eeq
   as in Fig. \ref{cerocero} in which we plot an example with a dS maximum. This quantity should be either negative or larger than  a certain constant $c$. Since there is a dS maximum at $T=\rho$, it will not be true at this point for any $c$ and for other points after fixing $c$. According to \cite{Ooguri:2018wrx} for those points we have to make sure that the smallest eigenvalue of the Hessian satisfies: 
\begin{equation}
\text{min}\left(\nabla_{i}\nabla_{j}V\right)\leq-c'V
\end{equation} 
\begin{figure}[H]
	\centering{}
	\includegraphics[scale=0.39]{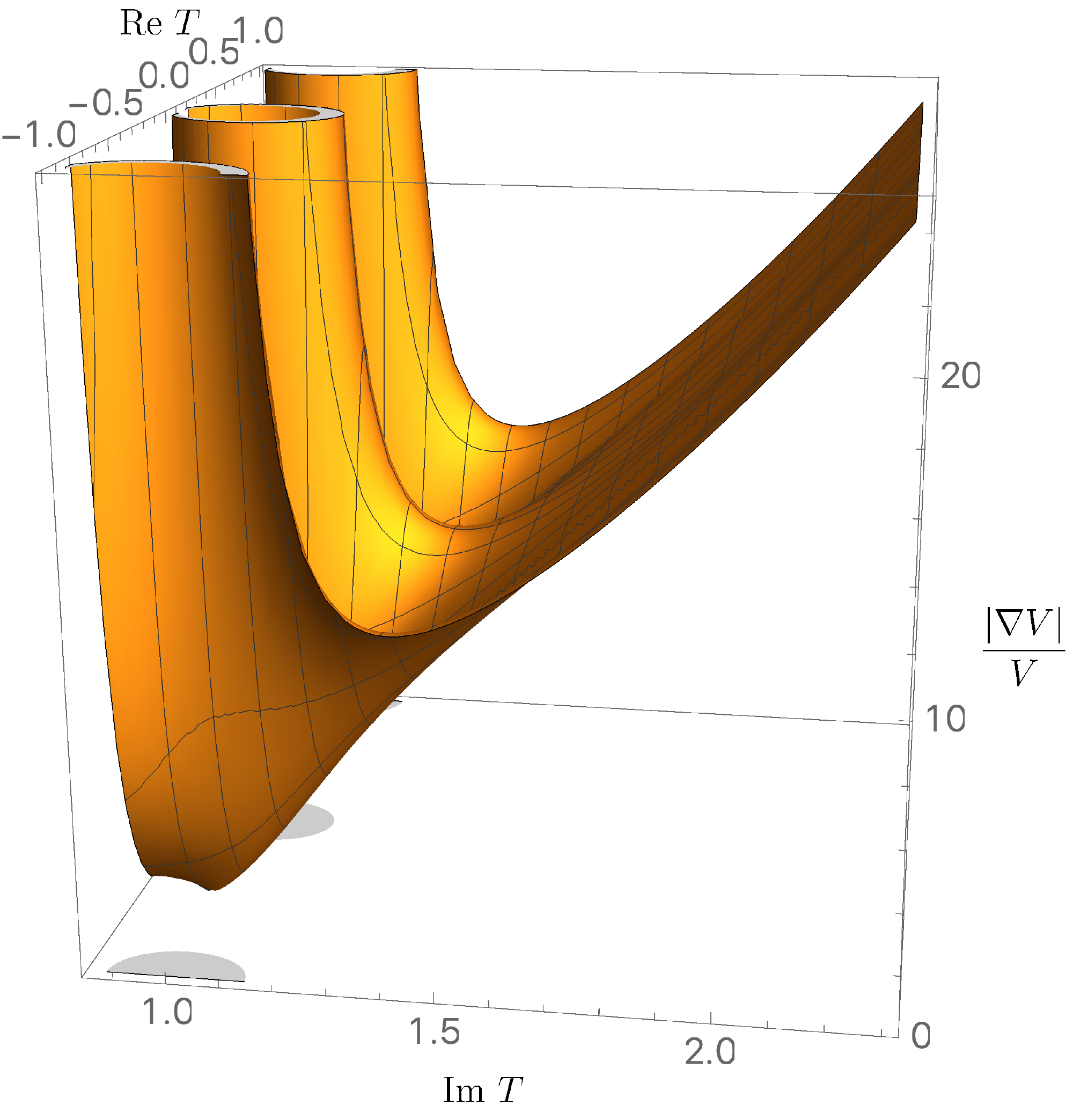}
	\caption{\footnotesize $\frac{\left|\nabla V\right|}{V}$ for $W=j^{1/3}/\eta^6$. The ratio is always bigger than one and grows linearly with $\text{Im} \ T$.}
	\label{cerocero}
\end{figure}
In all examples analyzed with a dS maximum, the third condition on the Hessian is satisfied.
One can also check whether the dS condition is obeyed for large moduli, away from any dS maxima, which in fact can be studied analytically.
 Taking $j\rightarrow e^{-i2\pi T}$, $\hat{G}_{2}(T)\rightarrow\frac{\pi^{2}}{3}$ and $H\rightarrow e^{2\pi\left(\frac{m}{2}+\frac{n}{3}+r\right)T}{\cal P}\left(e^{-i2\pi T}\right)$ in Eq. (2.20) of \cite{Cvetic:1991qm} we find:
\begin{equation}
\frac{\left|\nabla V\right|}{V}=\frac{2}{\sqrt{3}}T_{\text{I}}\pi \left(\frac{6m+4n+2r+3}{6}\right),
\end{equation}
where $r$ is the highest exponent of the polynomial $\cal P$. This quantity is always positive and it goes to infinity for large $T$, so it is guaranteed that in this limit the dS conjecture holds. Again, we resorted to numerical methods to check that it holds everywhere in all examples.

It is interesting to note that this is an specific example of a large class of scalar potentials which obey the refined swampland conjecture but
do not correspond to a simple monotonous decreasing potential, as considered up to now, and leading to quintessence type of potentials.

We have not much  to say about the AdS conjecture of ref.\cite{OV}  in these one-modulus models.  This conjecture states that there are  no
non-SUSY, AdS stable vacua in a theory consistent with quantum gravity.  In the class of models described above all AdS vacua in the 
self-dual points are SUSY. There are non-SUSY AdS vacua  only in other points at the boundary of the fundamental region, as in the 
simple example with $n=m=0$, ${\cal P}=1$ depicted in fig.(2). Such vacua should be unstable if the conjecture is correct,
perhaps decaying by a mechanism analogous to that of bubbles of nothing, as in \cite{Witten2}. It would be interesting to check if such an instability exists.

\subsection{Two moduli model}

The above single-modulus class of potentials is already very rich. However generic string compactifications yield 
effective field theories which contain typically multiple moduli, gauge groups and charged particles. In the end the total vacuum energy
depends on multiple contributions and a subsector which by itself would yield e.g. an AdS vacuum, may be perhaps overwhelmed by
other sectors of the theory, yielding positive energy and a  dS minima. This is something that we cannot predict on the basis of the 
potential of a subsector of the theory.  

In this sense, it is interesting to check whether the results found above for a single modulus is very much modified by the addition of 
an extra chiral multiplets $S$.  A simple example is provided by the heterotic gaugino condensation superpotential  discussed in section
3.
So, inspired by the gaugino condensate model one  can consider an $\NN=1$ supergravity model
\begin{equation}
G \left(S,S^{*},T,T^{*}\right)=  -\log \left(S-S^{*}\right)-3 \log{ \left(T-T^{*}\right)}\ +\log|W(S,T)|^2
\end{equation} 
with
\begin{equation}
W\left(S,T\right)=\frac{\Omega\left(S\right)H\left(T\right)}{\eta\left(T\right)^{6}}.
\end{equation}
The resulting scalar potential is then given by:
\beqa
&& \!\!\!\! \!\!\!\! \!\!\!\!  V\left(T,T^{*},S,S^{*}\right)  =\\
&& =\, \frac{1}{16S_{I}T_{I}^{3}\left|\eta\right|^{12}}\biggl\{\left|2 i S_{I}\Omega_{S}-\Omega\right|^{2}\left|H\right|^{2}+\frac{4T_{I}^{2}}{3}\left|\frac{dH}{dT}+\frac{3}{2\pi}H\hat{G}_{2}\right|^{2}\left|\Omega\right|^{2} -3\left|H\right|^{2}\left|\Omega\right|^{2}\biggr\} \ . \nonumber
\eeqa
We denote derivatives with respect to the fields with a sub-index $T$ or $S$. 
In  Appendix B we study the extrema of this general potential in some detail. 
For general $\Omega$ there are essentially two classes of vacua, depending on whether the auxiliary field of $S$, $F_S$ vanishes or not at the minimum. If it vanishes,
the structure of vacua are identical to the single modulus case discussed above and the same conclusions apply.  For $F_S\not=0$ there are new possibilities
which are summarized in the tables  in  Appendix B.  Essentially the same classes of minima as in the single modulus case are obtained and again
no dS minima are found.

\section{Conclusions and outlook}
\label{sec:conclu}

In this paper we have put forward the use of the modular symmetries in the effective action for moduli to test swampland conjectures in regimes beyond weak coupling expansion points. On one hand, we have constructed explicit string theory compactifications in which such modular symmetries are ingrained in the UV construction, and derived how several swampland constraints arise from the dynamics of e.g. non-perturbative superpotentials in models with 4d $\NN=1$ supersymmetry. On the other hand, we have discussed fairly general effective theories for scalars enjoying duality modular symmetry and explored the extent to which the latter imply satisfying the swampland constraints.

The string constructions we have considered are explicit toroidal orbifolds of the heterotic string, for which moduli in $\NN=2$ subsectors from fixed tori have non-trivial dynamics due to their appearance in threshold corrections of 4d gauge kinetic functions. In particular, we have described examples in which the access to points at infinity in moduli space, where a global shift symmetry for the axion would be restored, is dynamically forbidden by exponential growth of the (non-perturbative) scalar potential. This potential arises from infinite towers of states becoming light, so the absence of global symmetries is interestingly linked to the swampland distance conjecture in a very explicit and novel mechanism.
This new mechanism may well be the counterpart, in the context of lower (or no) supersymmetry, of the mechanisms discussed in the literature in cases with 8 supercharges.

We have also considered several dual pictures, in particular type I' / Horava-Witten theory compactifications, in which the degrees of freedom associated to complex structure or K\"ahler moduli can be studied in isolation. Such states are key to the derivation of the swampland distance conjecture in the present context. We expect that, given the ubiquitous appearance of modular groups in CY moduli spaces, and of string dualities along the lines discussed, the strategy of relating modular properties with consistency with quantum gravity may apply in a far more general context, which we hope to explore in the future.

We have performed a careful numerical analysis of the properties of scalar potentials in general effective theories for scalars, with essentially the only constraint that they enjoy invariance under modular symmetries. Surprisingly, this requirement alone seems to suffice to render the theories consistent with a number of swampland conjecture, besides the 
already mentioned distance conjectures. The potentials have extrema that correspond to (supersymmetric) Minkowski and AdS minima, and dS maxima or saddle points, but no dS minima. Moreover the potentials satisfy the refined de Sitter conjecture conditions. These  remarkable facts  suggest the tantalizing proposal that modular properties are deeply ingrained in the conditions to guarantee consistency of effective theories with quantum gravity. This is certainly a promising research direction to pursue.  
Let us however insist that we do not claim that the fact that we have been unable to find any dS minimum should apply to any consistent theory of quantum gravity. 
Only that in the simplest classes of modular invariant effective actions we have been unable to find any dS minimum.

A final remark concerns the fact that, naively, modular symmetries are manifest in the effective actions of compactifications in which further ingredients are absent. For instance, introduction of fluxes typically induces scalar potentials which are not  invariant under the modular transformations, thus seemingly breaking the symmetry explicitly at the level of the effective action. In fact, this interpretation is far from complete, and there is a precise sense in which the symmetry is present. The situation is analogous to the statement that e.g. in heterotic compactifications on K3$\times \IT^2$, the presence of Wilson lines on $\IT^2$ would seem to break the $SL(2,\IZ)$ symmetries, since the Wilson lines are not invariant. In fact, the correct interpretation is that the duality symmetries act non-trivially on the Wilson lines  and define a larger duality group $SO(4,20;\IZ)$ acting on and enlarged moduli space which includes the Wilson lines. Similarly, in flux compactifications, the duality groups act consistently on the set of all flux vacua by acting non-trivially on both the moduli and the fluxes. Restricting to axion scalars, this is just a re-statement that the flux potentials have an axion monodromy structure. The extension to the full duality group brings in a much richer set of fluxes, necessarily including non-geometric ones. We expect that the constraints of invariance under modular transformations of moduli in the landscape of flux vacua bring new insights into both the structure of the landscape and the set of conditions discriminating it from the swampland.

\vspace{2.0cm}

\centerline{\bf \large Acknowledgments}

\vspace{0.3cm}

\noindent We thank A. Font, A. Herr\'aez,  F. Marchesano, M. Montero and I. Valenzuela for useful discussions. 
This work has been supported by the ERC Advanced Grant SPLE under contract ERC-2012-ADG-20120216-320421, by the grants FPA2016-78645-P and
FPA2015-65480-P from the MINECO
 and the grant SEV-2016-0597 of the ``Centro de Excelencia Severo Ochoa" Programme.  The work of E. Gonzalo is supported by a FPU fellowship 
 number FPU16/03985.

\newpage

\appendix

\section{ $SL(2,{\bf Z})$ modular forms}
\label{sec:appa}

Here we just list some well known properties for the modular forms discussed in the main text. We follow closely the
appendix in \cite{Cvetic:1991qm}. The modular group is generated by
the transformations
\beq
T\ \longrightarrow \ \frac {aT+b}{cT+d} \ ,  \ a,b,c,d\in {\bf Z}, ad-bc=1 \ ,
\eeq
with $T=\theta+it$. This is is generated by a shift transformation $T\rightarrow T+1$ and $T\rightarrow -1/T$ which relates small and large $t$.
Since changing the sign of all the parameters leaves invariant the transformation, the group is actually $PSL(2,{\bf Z})=SL(2,{\bf Z})/{\bf Z_2}$. 
A meromorphic function $F(T)$ is
said to have modular weight $r$ if
\beq
F\left(\frac {aT+b}{cT+d}\right) \ =\ (cT+d)^rF(T) \ .
\eeq

The symmetry 
divides the complex plane in equivalent regions and the conventional fundamental domain ${\cal D}$ is shown in fig.(1).
 The group $PSL(2,{\bf Z})$ has self-dual points  at $T=i,\rho$ with $\rho=e^{i2\pi/3}$, as well as at  infinity. A modular form admits a Laurent expansion 
 at each point in the 
 interior of the closed domain ${\cal D}$. The lowest order of the expansion of  $F$ in $p$ is denoted $\nu_p$.
 Around the self-dual points $F$ has an expansion
 in terms of uniformizing variables. In particular, as $T\rightarrow \infty$ one expands
 \beq
 F(T)\ =\ q^{\nu_\infty}\sum_{n=0}^\infty a_nq^ n \ , \  q\ =\ e^{i2\pi T}
 \eeq
 and $\nu_\infty$ is the order of $F$ at infinity. At $T=i,\rho$ the uniformizing variables are 
 \beq 
 t_1\ =\ \left(\frac {(T-i)}{(T+i)}\right)^2 \ \ , \ \ 
 t_\rho\ =\ \left(\frac {(T-\rho)}{(T+\rho^*)}\right)^3 \ .
 \eeq
In terms of them $F(T)$ admits expansions around $T=1,\rho$ of the form
\beq
(T+i)F(T)\ =\ t_1^{\nu_1/2}\ \sum_{n=0}^\infty \ b_nt_1^n \ ,\ 
(T+\rho)F(T)\ =\ t_\rho^{\nu_\rho/2}\ \sum_{n=0}^\infty  \ c_nt_1^n  \ .
\eeq
It may be proven that the orders are related to the modular weight $r$ by the expression
\beq 
\frac {r}{12} \ =\ \nu_\infty\ +\ \nu_1 \ +\ \nu_\rho \ + \sum_{p\not= 1,\rho,\infty} \nu_p \ .
\eeq
One interesting consequence of this formula  is that for negative modular weight $r<0$ (as in the physics examples in the
main tex) there must necessarily be a singularity in the fundamental domain.  Thus if, on physics grounds, there are no such singularities,
then $\nu_\infty<0$ and the exponential growth of $F(T)$ for large $ \text{Im} \ T$ is implied.

Of particular interest are the modular forms constructed from the Eisenstein series
\beq
G_{2k}(T)\ =\ \sum_{n_1,n_2\in {\bf Z}} \frac {1}{\left(n_1T\ +\ n_2\right)^{2k}} \ .
\eeq
For $k>1$ these are holomorphic modular forms of weight $2k$.  For $k=1$ $G_2$ rather transforms as a connection
$G_2\rightarrow (cT+d)^2G_2$$-2\pi c(cT+d)$. Then the the non-holomorphic
${\hat G}_2$ modular form
\beq
{\hat G}_2(T,T^*)\ =\ G_2(T)\ -\ \frac {\pi}{\text{Im} \ T}
\eeq
transforms as a modular form of weight 2. There is a {\it cusp form} of weight 12 given by
\beq 
\Delta(T) \ =\ \frac {675}{256\pi^{12}} \left[20G_4^3\ -\ 49G_6^2\right], 
\eeq
The Dedekind function is given by
\beq
\eta(T)\ =\ \Delta(T)^{1/24} \ .
\eeq
One can check the identity, used in the main text
\beq
\frac {d \eta(T)}{dT}\ =\ -\frac {1}{4\pi}\eta(T)G_2(T) \ .
\eeq
The holomorphic Klein modular invariant form $j(T)$ may be written in terms of the cusp form as
\beq
j(T)\ =\ \frac {91125}{\pi^{12}} \frac {G_4(T)^3}{\Delta(T)}  \ .
\eeq
It may be shown that any holomorphic modular invariant form is a rational function of $j(T)$.
Any modular form of weight $r$ which is regular in ${\cal D}$ may be written as
\beq
H(T)\ =\ (j(T)-1728)^{m/2}j(T)^{n/3}{\cal P}(j(T))
\eeq
and equivalently as
\beq
F(T)\ =\ \eta(T)^{2r}\left( \frac {G_6(T)}{\eta(T)^{12}}\right)^m\left(\frac {G_4(T)}{\eta(T)^8}\right)^n {\cal P}(j(T)) \ ,
\eeq
with $n,m$ positive integers and ${\cal P}$ a polynomial of $j(T)$. Some useful expansions in powers of $q=e^{i2\pi T}$ are
\beq
\eta(T)^{-1}\ =\  q^{-1/24}\left(1\ +\ q\ +\ 2q^2\ +\ 3q^3+...\right)
\eeq
\beq
G_2(T)\ =\ \frac {\pi^2}{3}\left(1\ -\ 24q\ -\ 72q^2-96q^3\ -...\right)
\eeq
\beq
j(T)\ =
 \frac {1}{q}\ +\ 744\ +\ 196884q\ +\ 21493760q^2\ +\ ...  \  .
 \eeq
 The divergence of $\eta(T)^{-1}$ at large $t$ is at the origin of the superpotential divergences for large modulus.

\section{General Analysis of Minima for two fields}
\label{sec:appb}

In this Appendix we study the extrema of the potential in the two moduli model inspired by gaugino condensation. Let us recall from \cite{Cvetic:1991qm} that the extrema from Section 4 arise as a particular case. The second field provides a new condition for spontaneously susy breaking:
\begin{equation}
h^{S}\propto\left|H\right|\left(2 i S_{I}\Omega_{S}-\Omega\right)\not= \ 0
\end{equation}
Using this equation we can distinguish two types of extrema. In Type B extrema SUSY is broken spontaneously by the minimization in S. In contrast, in Type A extrema, SUSY remains unbroken 
 ($h^{S}= 0$)  at this stage. The condition $\frac{\partial V}{\partial S}=0$ for Type A extrema reads:
\begin{equation}
2 i S_{I} \Omega_{S}- \Omega = 0 \label{eq:susy_minima},
\end{equation}
while for Type B:
\begin{equation}
\frac{ -4 S_{I}^{2} \Omega_{SS} e^{-2i  \left| 2 i S_{I} \Omega_{S}- \Omega \right| } } {\Omega^{*}} = 2- 4 \frac{T_{I}^{2}} {3\left| H\right|^{2}} \left| \frac{dH}{dT}+\frac{3}{2 \pi} H \hat{G}_{2}\right|^{2}.
\end{equation}
If the S field is in a Type A extrema,  the dependence on $S$ after its minimization is trivial, since the potential is the one we would find for a single modulus $T$ multiplied by the constant $ \frac{ | \Omega(S) |^{2}}{2S_{I}}$. In this way, we recover the single modulus potential and  the results of Section 4 which we discussed in the main text. Even tough in
Type B extrema the dependence on $ \Omega (S)$ is not trivial anymore, we will see how the results can be easily extended. In particular the proof that there are no dS minima in the self-dual points or in the zeros of $\cal P$ can also be done analytically, except for the $n=0$ or $m=0$ case, for which we rely on numerics. Now we go through the same steps of Section 4 to show that if at the self-dual points there is a minimum then the potential cannot have positive minima.

\subsubsection*{$n>1$ or $m>1$}
If $n>1$ then at $T=\rho$ $H=\frac{dj}{dT}=0$, and the same happens at $T=i$ for $m>1$. The the potential is simply $V=0$ in both extrema. Just like it happened for Type A, if $n>1$ or $m>1$ then $T=\rho$ or $T=i$ respectively are always minima. 

\subsubsection*{$n=1$ or $m=1$}
If $n=1$ then $\frac{dH}{dT} \neq 0$ so the condition for Type B extrema reduces to $\Omega=0$. Therefore $V=0$ at both extrema. By changing H we can make it a saddle point or a maximum but not a minimum.

\subsubsection*{$n=0$ or $m=0$}
This case is the exception, for which analytically we have not been able to prove our result. At $T=\rho$: $\hat{G}_{2}=\frac{dj}{dT}=0$ so if $n=0$ then $\frac{dH}{dT}=0$ and
\begin{equation}
V\left(T,T^{*}\right) = \frac{\left|  {\cal P}\left( 0 \right)  \right|^{2}}{16 S_{I} T_{I}^{3}\left|\eta\right|^{12}}\left\{ \left|2 i S_{I}\Omega_{S}-\Omega\right|^{2}-3\left|\Omega\right|^{2}\right\} 
\end{equation}
It is a minimum if $\left|2 i S_{I}\Omega_{S}-\Omega\right|^{2}-2\left|\Omega\right|^{2}>0$ (only possible in extrema of Type B) and a maximum if $\left|2 i S_{I}\Omega_{S}-\Omega\right|^{2}-2\left|\Omega\right|^{2}<0$. There is no simple way to decide the sign of the vacuum energy  in this case. If we consider $\Omega$ as a sum of exponentials numerically we find that it is always negative.
At $T=i$: $j=1728$, $\hat{G}_{2}=\frac{dj}{dT}=0$, $\frac{dj}{dT}=0$, $ \left| H \right|=1728^{m/2} \left| {\cal P}\left( 1728 \right)\right| $ so for $m=0$ $\frac{dH}{dT}=0$ and
\begin{equation}
V\left(T,T^{*}\right)= \frac{ \left| {\cal P}\left( 1728 \right) \right|^2}{16 S_{I} T_{I}^{3}\left|\eta\right|^{12}}\left\{ \left|2 i S_{I}\Omega_{S}-\Omega\right|^{2}-3\left|\Omega\right|^{2}\right\} 
\end{equation}
By choosing different $\cal P$ all types of extrema can occur at $T=i$: maximum, minimum or saddle point. Namely,  it is a maximum  if $-2.57<\frac{H'''}{H'}<-1.57$, a saddle point if $\frac{H'''}{H'}<-2.57$ or $\frac{H'''}{H'}>-1.57$ and a minimum if $\left|\frac{H''}{H}+1.19\right|>\frac{3}{2}$. Again we do not prove it analytically, but numerically we also find that $V<0$.

The remaining $H=0$ extrema are the multiple zeros of ${\cal P}\left(j\right)$. These are SUSY minima and are always in Minkowski.
We summarize the results for Type B extrema at the self-dual points in the next two tables.
\begin{table}[H]
	\begin{tabular}{|c|c|c|c|}
		\hline 
		& $V\left(T=\rho\right)$ & Type of Extrema & Susy \tabularnewline
		\hline 
		\hline 
		$n>1$ & $V=0$ & Minimum  & Yes \tabularnewline
		\hline 
		$n=1$ & $V=0$ & Min $\left|2 i S_{I}\Omega_{S}-\Omega\right|^{2}-\frac{2}{3}\left|\Omega\right|^{2}>0$  & No \tabularnewline
		&  & Max $\left|2 i S_{I}\Omega_{S}-\Omega\right|^{2}-\frac{2}{3}\left|\Omega\right|^{2}<0$ & \tabularnewline
		\hline 
		$n=0$ &  Numerically all min $V<0$  & Min $\left|2 i S_{I}\Omega_{S}-\Omega\right|^{2}-2\left|\Omega\right|^{2}>0$ & No \tabularnewline
		& $\propto\left\{ \left|2 i S_{I}\Omega_{S}-\Omega\right|^{2}-3\left|\Omega\right|^{2}\right\} $ & Max $\left|2 i S_{I}\Omega_{S}-\Omega\right|^{2}-2\left|\Omega\right|^{2}<0$ &  \tabularnewline
		\hline 

	\end{tabular}\caption{ Classification of Type B extrema at $T=\rho$.} \label{tabla3}
\end{table}

\begin{table}[H]
	\begin{tabular}{|c|c|c|c|}
		\hline 
		& $V\left(T=i\right)$ & Type of Extrema & Susy\tabularnewline
		\hline 
		\hline 
		$m>1$ & $V=0$ & Minimum & Yes \tabularnewline
		\hline 
		$m=1$ & $V=0$ & No simple criteria & No \tabularnewline
		&  &  & \tabularnewline
		\hline 
		$m=0$ & Numerically all min have $V<0$ & No simple criteria & No\tabularnewline
		&  $ \propto\left\{ \left|2 i S_{I}\Omega_{S}-\Omega\right|^{2}-3\left|\Omega\right|^{2}\right\} $  &  & \tabularnewline
		\hline 
	\end{tabular}\caption{Classification of Type B extrema at $T=i$.}		\label{tabla4}
\end{table}

 However, we have searched numerically for dS minima at other points on the boundary, supposing $\Omega$ is given by a sum of exponentials, for various choices of $H$ and we have not found any.


\end{document}